\documentclass[journal]{IEEEtran}
\usepackage[utf8]{inputenc} 
\usepackage[T1]{fontenc}    
\usepackage{graphicx} 
\usepackage{amsmath}
\usepackage{amsfonts}
\usepackage{caption}
\usepackage{subcaption}
\usepackage{algorithm}
\usepackage[noend]{algpseudocode}
\usepackage{makecell}
\usepackage{cite}
\usepackage{url}
\usepackage{multirow}
\usepackage{booktabs}

\usepackage[amsmath,thmmarks,framed]{ntheorem} 
\usepackage[style=0,ntheorem]{mdframed}
\mdfsetup{%
topline=false,
rightline=false,
bottomline=false,
linewidth=3pt,
innerleftmargin=15pt,
innerrightmargin=0pt,
skipabove=\baselineskip,
skipabove=1.2\baselineskip,
}
\newtheorem{thm}{Theorem}
\newtheorem{definition}{Definition}
\newtheorem{col}{Corollary}
\newtheorem{lemma}{Lemma}
\begin{document}
\title{On Purely Data-Driven Massive MIMO Detectors}
\author{Hao Ye, \IEEEmembership{Member,~IEEE}  and Le Liang, \IEEEmembership{Member,~IEEE}
\thanks{H. Ye is with the Department of Electrical and Computer Engineering, the University of California, Santa Cruz, CA, USA. (e-mail: yehao@ucsc.edu)}
\thanks{L. Liang is with the National Mobile Communications Research Laboratory, Southeast University, Nanjing 210096, China (e-mail: lliang@seu.edu.cn). }
}
\maketitle
\begin{abstract}
    
    The development of learning-based detectors for massive multi-input multi-output (MIMO) systems has been hindered by the inherent complexities arising from the problem's high dimensionality. 
    To enhance scalability, most previous studies have adopted model-driven methodologies that integrate deep neural networks (DNNs) within existing iterative detection frameworks.
    However, these methods often lack flexibility and involve substantial computational complexity.
    In this paper, we introduce ChannelNet, a purely data-driven learning-based massive MIMO detector that overcomes these limitations.
    ChannelNet exploits the inherent symmetry of MIMO systems by incorporating channel-embedded layers and antenna-wise shared feature processors.
    These modules maintain equivariance to antenna permutations and enable ChannelNet to scale efficiently to large numbers of antennas and high modulation orders with low computational complexity, specifically $\mathcal{O}(N_t N_r)$, where $N_t$ and $N_r$ denote the numbers of transmit and receive antennas, respectively.
    Theoretically, ChannelNet can approximate any continuous permutation-symmetric function and the optimal maximum likelihood detection (ML) function with arbitrary precision under any continuous channel distribution.   
    Empirical evaluations demonstrate that ChannelNet consistently outperforms or matches state-of-the-art detectors across different numbers of antennas, modulation schemes, and channel distributions, all while significantly reducing computational overhead. 
    This study highlights the potential of purely data-driven designs in advancing efficient and scalable detectors for massive MIMO systems.

\end{abstract}

\section{Introduction}
Massive multi-input multi-output (MIMO) leverages a large number of antennas to improve data rates, coverage, and overall network performance. 
By facilitating simultaneous communication of multiple users, massive MIMO dramatically improves spectral efficiency and network capacity, which are essential for meeting the escalating demand for high-speed connectivity in the 5G era and beyond \cite{larsson2014massive, lu2014overview, chataut2020massive}. 
Despite its potential, the increased complexity associated with massive MIMO systems presents a significant barrier to fully realizing these benefits. 
Especially, the combination of large numbers of antennas and high-order modulations significantly heightens the challenge of efficient signal detection, rendering exact maximum likelihood (ML) detection computationally infeasible. 
For a system with $N_t$ transmit antennas and a modulation of $M$ symbols, the ML detector has an exponential complexity of $\mathcal{O}(M^{N_t})$.
Consequently, crafting an efficient and low-complexity symbol detector is essential to making massive MIMO practical for real-world applications.

There has been extensive research on designing efficient MIMO detectors \cite{yang2015fifty, larsson2009mimo}.
These approaches can be broadly categorized as linear, iterative, and learning-based detectors.
While linear detectors, such as zero-forcing (ZF) and minimum mean-squared error (MMSE), have low complexity and perform adequately in small-scale systems, they exhibit poor performance compared to the ML detector when large numbers of antennas and high-order modulations are utilized. 
The iterative detectors, such as approximate message passing (AMP) \cite{jeon2015optimality}, orthogonal AMP (OAMP) \cite{ma2017orthogonal},  and expectation propagation \cite{cespedes2014expectation}, aim to approximate the ML performance through iterative processes.
For example, the AMP detector is asymptotically optimal for large MIMO systems with independent and identically distributed (\textit{i.i.d.}) Gaussian channels but degrades significantly under realistic ill-conditioned channels \cite{jeon2015optimality}.
Subsequent research has focused on relaxing this \textit{i.i.d.} Gaussian assumption \cite{ma2017orthogonal, guo2015approximate, luo2021unitary}.
Orthogonal AMP (OAMP), for instance, achieves this improved robustness but requires more extensive computations due to the need for inverse matrix calculations in each iteration  \cite{ma2017orthogonal}.

Recent advancements in MIMO detection have increasingly emphasized learning-based approaches to leverage the expressive power of deep learning.
Early studies explored purely data-driven frameworks, where off-the-shelf deep neural networks (DNNs) were trained for predicting the transmitted symbols based on the channel matrix and the received signal. 
While effective for small-scale MIMO and low-order modulations (\textit{e.g.}, $4 \times 4$ MIMO with quadrature phase shift keying (QPSK) modulation \cite{hu2023understanding}), these methods struggle to scale to massive MIMO scenarios with higher-order modulations.
To address these limitations, recent efforts have shifted towards model-driven approaches that integrate DNNs with existing iterative detection frameworks \cite{he2020model, khani2020adaptive, goutay2020deep, yang2022deep, he2023message, yu2023data} or optimization algorithms \cite{zilberstein2022annealed, samuel2017deep, pratik2020re}. Despite their advancements, these methods heavily rely on the original frameworks, which often leads to high computational demands and reliance on specific assumptions about channel and noise characteristics, potentially compromising their effectiveness in real-world applications. Deep learning techniques have also assisted sphere decoding algorithms \cite{liao2022deep, weon2020learning, yue2024learned}, improving search efficiency. However, due to the inherent complexity of sphere decoding, these learning-assisted methods have mainly been applied effectively to small-scale MIMO systems.

In this paper, we revisit the purely data-driven paradigm and introduce ChannelNet, a novel architecture designed to address the scalability bottlenecks of earlier data-driven approaches.
ChannelNet employs off-the-shelf neural architectures to iteratively update and exchange features representing the transmit and receive antennas.
During each iteration, the multi-layer perceptron (MLP) models update the antenna features in an identical and independent manner across antennas.
Feature exchanges are facilitated through linear layers that directly incorporate the channel matrix. 
These designs enforce invariance and equivariance to antenna permutations, effectively reducing the search space and enabling efficient learning in large-scale massive MIMO systems.

The effectiveness of ChannelNet is supported through both theoretical analysis and empirical evaluation. 
As a purely data-driven solution, ChannelNet is able to unleash the expressive power of DNNs for function approximation.
ChannelNet can approximate any continuous permutation-symmetric function and the optimal ML detection function with arbitrary precision under any continuous channel distribution.  
Empirical results demonstrate that ChannelNet consistently matches or outperforms state-of-the-art detectors across different numbers of antennas, modulation schemes, and channel distributions. 
Moreover, ChannelNet exhibits enhanced robustness in novel environments compared to the model-driven detectors.

In addition to its superior performance, ChannelNet is also computationally efficient. With a complexity of $\mathcal{O}(N_t N_r)$, where $N_t$ and $N_r$ denote the number of transmit and receive antennas, respectively, ChannelNet significantly reduces the computational overhead compared to state-of-the-art learning-based detectors.  Furthermore, unlike most existing detectors that require noise power estimation, ChannelNet operates without this prerequisite, which simplifies implementation and alleviates the need for explicit noise power estimation.

In summary, our contributions can be outlined as follows:
\begin{itemize}
    \item We demonstrate the feasibility of purely data-driven massive MIMO detectors by presenting ChannelNet, which utilizes off-the-shelf neural network components and overcomes the scalability bottleneck without the reliance on predefined detection frameworks.
    \item We establish that ChannelNet can approximate any continuous permutation-symmetric function and the optimal ML detection function with arbitrary precision under any continuous channel distribution. 
    \item We conduct extensive experiments to empirically validate ChannelNet's effectiveness and efficiency across a wide range of scenarios, including varying numbers of antennas, modulation orders, and channel distributions. 
\end{itemize}

The remainder of this paper is organized as follows. Section  \ref{sec:preliminary} introduces the necessary preliminaries. In Section \ref{sec:methodology}, we detail ChannelNet, the proposed data-driven massive MIMO detector.   Section \ref{sec:theory} provides a theoretical analysis of ChannelNet's expressive power. The experimental results are presented in Section \ref{sec:experiment}, followed by conclusions in Section \ref{sec:conclusion}.

\section{Preliminaries}\label{sec:preliminary}

\subsection{Notation}
We adopt the following notation conventions: lowercase letters for scalars, bold lowercase letters for column vectors, and bold uppercase letters for matrices.
The $i$-th element of vector $\mathbf{x}$ is denoted as $\mathbf{x}_i$. The element in the $i$-th row and $j$-th column of matrix $\mathbf{H}$ is represented by $\mathbf{H}_{ij}$.
Unless mentioned otherwise, the superscript $t$ attached to functions, vectors, and matrices indicates the iteration step.
The symbol $\mathcal{S}_n$ denotes all possible permutations of $n$ elements, and $\mathbf{I}_n$ denotes the identity matrix of size $n$.

\subsection{Massive MIMO Detection Problem}
Consider an uplink scenario in which a massive MIMO base station with $N_r$ antennas serves $N_t$ single-antenna users, where $N_t \leq N_r$.
We assume a frequency-flat channel, and the channel coefficients between the $N_t$ transmit antennas and the $N_r$ receive antennas are represented by the matrix $\mathbf{\tilde{H}} \in \mathbb{C}^{N_r \times N_t}$, where $\mathbf{\tilde{H}}_{ij}$ denotes the channel from the $j$-th transmit antenna to the $i$-th receive antenna. 
The $N_t$ users transmit their symbols individually, forming a symbol vector $\mathbf{\tilde{x}} \in \mathbb{C}^{N_t}$, with each symbol belonging to a constellation $\mathcal{\tilde{X}}$. In this paper, we primarily focus on square quadrature amplitude modulation (QAM), and the symbols are normalized to attain unit average power. It is assumed that the constellation is the same for all transmitters, and each symbol has the same probability of being chosen by the users.  The received vector $\mathbf{\tilde{y}} \in \mathbb{C}^{N_r}$ is formed as the transmitted symbol vector propagates through the channel $\mathbf{\tilde{H}}$ and is corrupted by additive noise $\mathbf{\tilde{n}}$. This relationship is expressed as:
\begin{equation}
    \mathbf{\tilde{y}} = \mathbf{\tilde{H}\tilde{x}} + \mathbf{\tilde{n}}.
\end{equation}
In this study, we adopt an equivalent real-valued representation by separately considering the real $\Re(.)$ and imaginary $\Im(.)$ parts. Let $\mathbf{x} = [\Re(\mathbf{\tilde{x}})^T, \Im(\mathbf{\tilde{x}})^T]^T \in \mathbb{R}^K$, $\mathbf{y} = [\Re(\mathbf{\tilde{y}})^T, \Im(\mathbf{\tilde{y}})^T]^T \in \mathbb{R}^N$, $\mathbf{n} = [\Re(\mathbf{\tilde{n}})^T, \Im(\mathbf{\tilde{n}})^T]^T \in \mathbb{R}^N$, and
\begin{equation}
\mathbf{H} =
\begin{bmatrix}
\Re(\mathbf{\tilde{H}}), &-\Im(\mathbf{\tilde{H}}) \\
\Im(\mathbf{\tilde{H}}), & \Re(\mathbf{\tilde{H}})
\end{bmatrix} \in \mathbb{R}^{N \times K},
\end{equation}
where $K = 2N_t$ and $N = 2N_r$. This allows us to express the system in the equivalent real-valued form as follows:
\begin{equation}
\mathbf{y} = \mathbf{Hx + n}.
\end{equation}

Massive MIMO detectors aim to determine the transmitted vector $\mathbf{x}$ in a real-valued constellation $\mathcal{X}^K$ based on the received vector $\mathbf{y}$, and the channel matrix $\mathbf{H}$. The channel matrix is typically assumed to be known at the receiver but not at the transmitter.\footnote{The channel matrix $\mathbf{H}$ is typically obtained at the receiver through channel estimation techniques using pilot signals sent by the transmitter.} Formally, the massive MIMO detection problem involves solving the combinatorial optimization problem:
\begin{equation}
\mathbf{\hat{x}} = \operatorname*{argmin}_{\mathbf{x} \in \mathcal{X}^K} \| \mathbf{y - Hx}\|^2,
\end{equation}
which is NP-hard due to the finite constellation constraint. The ML detector serves as an optimal algorithm for solving the MIMO detection problem, employing an exhaustive search that assesses all possible symbols in the constellation. However, the complexity of the ML detector grows exponentially with the number of transmitted data streams, rendering it impractical for massive MIMO scenarios.

\subsection{Linear MIMO Detectors}

Linear detectors, such as ZF and MMSE detectors, are attractive due to their low computational complexity. However, their performance is not on par with the optimal ML detector. These linear detectors operate by multiplying the received signal $\mathbf{y}$ with an equalization matrix $\mathbf{A}$. Subsequently, the output undergoes quantization through a slicer denoted as $\mathcal{S(\cdot)}$, which quantizes each entry to the nearest neighbor in the constellation. The overall operation of linear detectors can be expressed as $\mathbf{\hat{x}} = \mathcal{S}(\mathbf{A} \mathbf{y})$.

\textbf{ZF detector:} The ZF detector neutralizes the effect of the channel matrix $\mathbf{H}$ by employing its pseudo-inverse as the equalization matrix, which is given by
\begin{equation}
    \mathbf{A_{ZF}} = (\mathbf{H}^T \mathbf{H})^{-1}\mathbf{H}^T.
\end{equation}

\textbf{Linear MMSE detector:} The linear MMSE detector improves upon the ZF method by accounting for noise, incorporating an additional term related to the noise power. The equalization matrix for the MMSE detector is expressed as
\begin{equation}
    \mathbf{A_{MMSE}} = ( \mathbf{H}^T \mathbf{H} + \sigma^2 \mathbf{I}_K )^{-1} \mathbf{H}^T,
\end{equation}
where $\sigma^2$ is the noise power. This approach effectively mitigates the noise enhancement problem inherent in the ZF detector, allowing the linear MMSE detector to achieve improved performance.

\subsection{Iterative MIMO Detectors}
Iterative MIMO detectors aim to approximate the ML solution by iteratively refining the estimates of the transmitted symbols. 
Many iterative detectors can be formulated within a general framework involving two steps per iteration:
\begin{equation}\label{equ:iter}
\begin{aligned}
    \mathbf{u}^{(t)} &= \mathbf{\hat{x}}^{(t)} + \mathbf{A}^{(t)}(\mathbf{y} - \mathbf{H\hat{x}}^{(t)}) + \mathbf{b}^{(t)}, \\
    \mathbf{\hat{x}}^{(t+1)} &= \eta^{(t)} (\mathbf{u}^{(t)}).
\end{aligned}
\end{equation}
The first step applies a linear transformation, yielding an intermediate signal $\mathbf{u}^{(t)}$ in a continuous domain. The second step employs a non-linear denoiser $\eta^{(t)}(\cdot)$, which processes $\mathbf{u}^{(t)}$ to produce the updated estimate $\mathbf{\hat{x}}^{(t+1)}$ for the next iteration. 

\textbf{Approximate message passing (AMP):}  AMP performs approximate inference on a bipartite graph representation of the MIMO system, achieving asymptotically optimal performance in large MIMO systems with \textit{i.i.d.} Gaussian channels \cite{jeon2015optimality}.  Despite its efficiency, its performance degrades significantly with other types of channels, especially ill-conditioned or spatially correlated channels, which are common in practical scenarios. AMP can be formulated as a specific case within the general framework \eqref{equ:iter} by setting $\mathbf{A}^{(t)} = \mathbf{H}^T$  and $\mathbf{b}^{(t)}$ as the Onsager term \cite{jeon2015optimality}.

\textbf{Orthogonal AMP (OAMP):} OAMP was introduced to extend AMP's applicability by relaxing the constraints on channel matrices, accommodating unitary-invariant matrices \cite{tulino2004random}. Also fitting within the general framework \eqref{equ:iter}, OAMP excludes $\mathbf{b}^{(t)}$ but involves a sophisticated $\mathbf{A}^{(t)}$, which requires computing a matrix pseudo-inverse at each iteration, resulting in a much higher complexity than the original AMP.

\subsection{Learning-Based MIMO Detectors}
Learning-based massive MIMO symbol detectors can be broadly classified into two paradigms: purely data-driven and model-driven approaches.\footnote{ The definitions of data-driven and model-driven detectors vary across different studies.  In this paper, we define model-driven detectors as those that incorporate predefined iterative algorithms into their design, while purely data-driven detectors operate independently of such frameworks.}
 Early purely data-driven approaches utilized off-the-shelf DNN architectures to predict the transmitted vector $\mathbf{x}$ by feeding the channel matrix $\mathbf{H}$ and the received signal $\mathbf{y}$ as input. While effective for small-scale MIMO systems with low-order modulation schemes (\textit{e.g.}, $4 \times 4$ MIMO with QPSK modulation \cite{hu2023understanding}), these methods struggle to scale to massive MIMO scenarios with higher-order modulations.
 Consequently, recent research has shifted focus towards model-driven architectures that integrate DNNs within predefined iterative detection or optimization algorithms. 
 
\textbf{DetNet:} DetNet is designed based on the iterative projected gradient descent algorithm \cite{samuel2017deep}. It performs well on \textit{i.i.d.} Gaussian channels with low-order modulations but exhibits performance degradation with higher-order modulations.

\textbf{OAMPNet:} OAMPNet unfolds the OAMP algorithm, introducing two trainable parameters per iteration to improve upon the original OAMP's performance \cite{he2020model}. 
However, similar to OAMP, OAMPNet requires computing a matrix pseudo-inverse in each iteration, resulting in higher complexity compared to alternatives such as AMP.

\textbf{AMP-GNN:} AMP-GNN unfolds the AMP algorithm and introduces a graph neural network (GNN) module for multi-user interference cancellation \cite{he2023message}. It achieves performance on par with state-of-the-art deep learning-based MIMO detectors while offering reduced computational complexity.

\textbf{MMNet:}  MMNet \cite{khani2020adaptive} builds upon the iterative soft-thresholding algorithm \cite{beck2009fast}. It can be formulated within the iteration framework \eqref{equ:iter} by considering $\mathbf{A}^{(t)}$ as a matrix of trainable parameters. For correlated channels, MMNet requires online training for each channel realization, thereby limiting its applicability.

\textbf{RE-MIMO:}  RE-MIMO \cite{pratik2020re} follows the recurrent inference machine framework \cite{putzky2017recurrent}, achieving state-of-the-art results for correlated channels. Each iteration consists of an encoder step, parameterized by a transformer, and a predictor module acting as the denoiser. 

\textbf{AMIC-Net:} AMIC-Net \cite{yu2023data}  unfolds a sequential detector \cite{mandloi2016low}, which iteratively detects transmitted symbols from each user while nullifying the interference from others.  It introduces a trainable extrapolation factor and a sparsely connected architecture to improve its performance.

Beyond the deep unfolding of iterative algorithms, some research has also focused on integrating deep learning with sphere decoding algorithms \cite{liao2022deep, weon2020learning, yue2024learned}. These learning-assisted sphere decoding methods typically use neural networks to help guide or simplify the search process.  However, due to its inherent complexity, these approaches remain practical mainly for small-scale MIMO systems.

Although these methods have shown improved performance, they are inherently tied to the assumptions and limitations of the underlying algorithms. This paper revisits the purely data-driven paradigm and proposes novel designs to overcome the scalability bottleneck of earlier data-driven methods.

\subsection{Permutation Invariant/Equivariant Neural Networks}
Permutation invariant and equivariant neural networks are specifically designed to handle unordered and symmetrically structured data by embedding symmetry properties within their architectures.  This incorporation enhances efficiency, robustness, and generalizability across various applications \cite{zaheer2017deep, maron2019universality, keriven2019universal}.
A function or neural network $f$ is considered permutation invariant if the output remains unchanged when the input is permuted.  Mathematically, for an input tuple $(x_1, x_2, \cdots, x_n)$ and any permutation $\pi \in \mathcal{S}_n$, this property can be expressed as:
\begin{equation}
    f(x_{\pi(1)}, x_{\pi(2)}, \cdots, x_{\pi(n)}) = f(x_1, x_2, \cdots, x_n).
\end{equation}
Similarly, a function or neural network $f$ is permutation equivariant if permuting the inputs results in the same permutation applied to the outputs:
\begin{equation}
     f(x_{\pi(1)}, x_{\pi(2)}, \cdots, x_{\pi(n)}) = \pi(f(x_1, x_2, \cdots, x_n)).
\end{equation}
Examples of permutation invariant networks include DeepSets \cite{zaheer2017deep} and PointNet \cite{qi2017pointnet}, while various types of GNNs are prominent examples of permutation equivariant networks. These architectures find applications in fields such as chemistry \cite{gilmer2017neural}, 3D point cloud processing \cite{qi2017pointnet}, and wireless resource allocation \cite{he2020resource}.

Beyond their practical success, significant research has been devoted to understanding their theoretical capabilities, particularly their expressive power. Standard DNNs, such as MLPs, are known as universal approximators, capable of approximating any continuous function with arbitrary precision \cite{cybenko1989approximation, hornik1989multilayer}. 
For permutation invariant and equivariant networks, universality pertains to their ability to approximate functions that adhere to the symmetries in the data. 
This property has been studied extensively for invariant networks \cite{zaheer2017deep, wagstaff2019limitations, wagstaff2022universal, maron2019universality, yarotsky2022universal} and equivariant networks \cite{keriven2019universal, sannai2019universal, dym2020universality}.

In massive MIMO detection, the channel matrix $\mathbf{H}$, received signal $\mathbf{y}$, and transmitted signal $\mathbf{x}$ are sensitive to antenna ordering. Permutations of receive antennas reorder entries in $\mathbf{y}$ and rows in $\mathbf{H}$, while permutations of transmit antennas reorder entries in $\mathbf{x}$ and columns in $\mathbf{H}$. Ideally, detectors should yield consistent results regardless of antenna permutations, being invariant to receive antenna permutations and equivariant to transmit antenna permutations.  Ignoring these symmetries significantly reduces the training efficiency for learning-based detectors as the model must relearn identical patterns with different antenna orderings. This study investigates data-driven detectors with architectures designed to preserve these symmetries—being invariant to receive antenna permutations and equivariant to transmit antenna permutations.  These symmetry-preserving designs effectively reduce the search space for learning, which is essential for improved scalability to large-scale massive MIMO systems.

\begin{figure*}[htbp]
    \centering
    \includegraphics[width=1\linewidth]{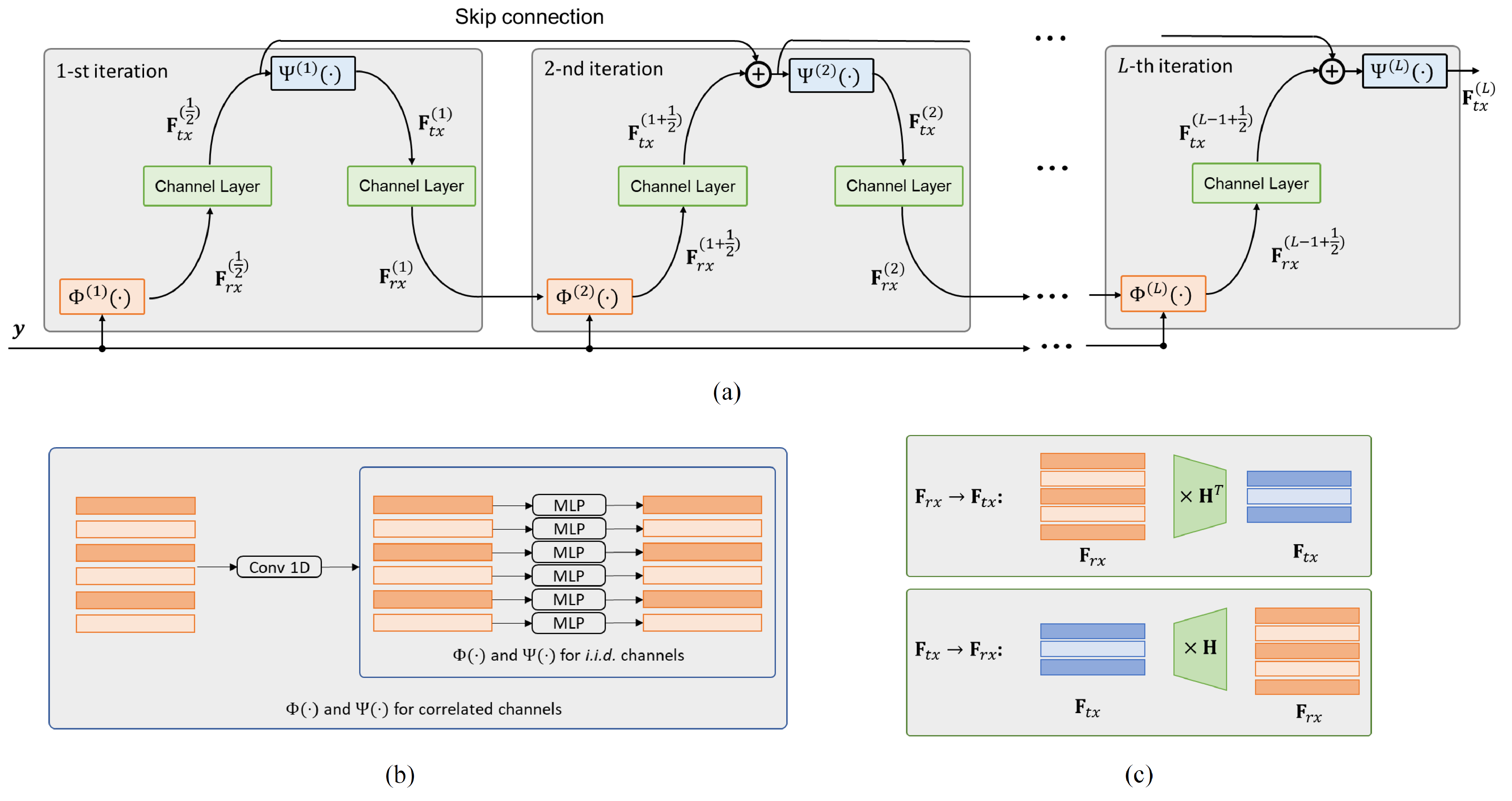}
    \caption{Proposed ChannelNet architecture: (a) The overall iterative framework of ChannelNet. (b) The antenna feature processor architectures tailored for \textit{i.i.d.} and correlated channels. (c) The channel layer architectures.}
    \label{fig:architecture}
\end{figure*}

\section{Method}\label{sec:methodology}

In this section, we provide a detailed description of the proposed ChannelNet, including its architecture, permutation symmetry, and connections to existing methods.

\subsection{Architecture}

ChannelNet employs neural networks to learn feature vectors representing transmit and receive antennas, which are iteratively refined and exchanged during the detection process. Specifically, we denote the transmit antenna feature as $\mathbf{F}_{\mathrm{tx}} = [\mathbf{f}_1^{\mathrm{tx}}, \ldots, \mathbf{f}_K^{\mathrm{tx}}] \in \mathbb{R}^{K \times d}$ and the receive antenna feature as $\mathbf{F}_{\mathrm{rx}} = [\mathbf{f}_1^{\mathrm{rx}}, \ldots, \mathbf{f}_N^{\mathrm{rx}}] \in \mathbb{R}^{N \times d}$, where $\mathbf{f}_i^{\mathrm{tx}}$ and $\mathbf{f}_j^{\mathrm{rx}}$ are $d$-dimensional feature vectors representing the $i$-th transmit and $j$-th receive antenna, respectively.

As shown in Fig. \ref{fig:architecture}a, the architecture of ChannelNet comprises two main modules: antenna feature processors and channel layers. The antenna processors update the  antenna features, while the channel layers enable structured transformations between transmit and receive antenna features. The process begins with the initialization of receive antenna feature $\mathbf{F}^{(0)}_{\mathrm{rx}}$ using the received signal $\mathbf{y}$. In the $t$-th iteration, the receive antenna feature $\mathbf{F}^{(t)}_{\mathrm{rx}}$ is first updated by the receive processor $\Phi^{(t)}(\cdot)$, resulting in an intermediate receive antenna feature $\mathbf{F}^{(t+\frac{1}{2})}_{\mathrm{rx}}$.
This intermediate receive antenna feature is passed through a channel layer, producing the corresponding intermediate transmit antenna feature $\mathbf{F}^{(t+\frac{1}{2})}_{\mathrm{tx}}$. Next, the transmit antenna feature is refined by the transmit processor $\Psi^{(t)}(\cdot)$, yielding $\mathbf{F}^{(t+1)}_{\mathrm{tx}}$. The updated transmit antenna feature is then passed through another channel layer to produce the receive antenna feature $\mathbf{F}^{(t+1)}_{\mathrm{rx}}$ for the next iteration. This sequence of alternating updates between the transmit and receive antenna features continues iteratively, gradually enhancing the feature representations with each pass. The iterative procedure concludes after $L$ iterations, producing the final transmit antenna feature $\mathbf{F}^{(L)}_{\mathrm{tx}}$ as the output.

\textbf{Antenna feature processor:}  
The antenna feature processors, $\Phi^{(t)}(\cdot)$ and $\Psi^{(t)}(\cdot)$, are trainable neural networks designed to update the feature vectors for transmit and receive antennas, respectively. 
As illustrated in Fig. \ref{fig:architecture}b, we use two different architectures for \textit{i.i.d.} channels and correlated channels.
For \textit{i.i.d.} channels, where channel distribution is independent of antenna ordering, we apply a local function paradigm \cite{zaheer2017deep, qi2017pointnet} to ensure permutation-equivariance. 
Specifically, each feature vector is independently updated using a shared MLP, ensuring that the updates are identical and do not depend on the antenna ordering.
Let $\phi^{(t)}(\cdot)$ and $\psi^{(t)}(\cdot)$ represent the MLPs used to update the feature vectors of receive and transmit antennas at the $t$-th iteration, respectively. The update process can be expressed as:
\begin{align}
    \Phi^{(t)}(\mathbf{F}_{\mathrm{rx}}) &= [\phi^{(t)}(\mathbf{f}^{\mathrm{rx}}_1),  \cdots,  \phi^{(t)}(\mathbf{f}^{\mathrm{rx}}_N)], \\
    \Psi^{(t)}(\mathbf{F}_{\mathrm{tx}}) &= [\psi^{(t)}(\mathbf{f}^{\mathrm{tx}}_1),  \cdots,  \psi^{(t)}(\mathbf{f}^{\mathrm{tx}}_K)].
\end{align}

For correlated channels, where the channel distribution depends on antenna ordering due to spatial correlations, convolutional layers are added before the MLPs in the antenna feature processors to capture dependencies between adjacent antennas, as shown in Fig. \ref{fig:architecture}b. This allows for improved feature updates by accounting for local antenna correlations.
In the following sections, the architectures of ChannelNet for \textit{i.i.d.} channels and correlated channels are referred to as ChannelNet-MLP and ChannelNet-Conv, respectively.

\textbf{Channel layer:} The exchanges between the transmit and receive antenna features are facilitated through the channel layers.
As depicted in Fig. \ref{fig:architecture}c, these layers utilize the channel matrix $\mathbf{H}$ as the coefficients of linear transformations between the transmit and receive antenna features.
Specifically, the transformation from transmit features to receive features is given by $\mathbf{F}_{\mathrm{rx}} = \mathbf{H} \mathbf{F}_{\mathrm{tx}}$. 
Conversely, the transformation from receive features to transmit features is given by $\mathbf{F}_{\mathrm{tx}} = \mathbf{H}^T \mathbf{F}_{\mathrm{rx}}$. 
By embedding the channel matrix $\mathbf{H}$ directly as linear layers within the network, the need to explicitly input $\mathbf{H}$ is eliminated, simplifying the architecture and enhancing the network's ability to efficiently model the interactions between transmit and receive features.

\begin{algorithm} 
\caption{ChannelNet Detector}\label{alg:main}
\begin{algorithmic}[1]
\Procedure{ChannelNet}{$\mathbf{H, y}$} 
\State $\mathbf{F}_{\mathrm{rx}} \gets \mathbf{y}$
\For{$t = 1$ to $L$}
    \State $\mathbf{F}_{\mathrm{rx}} \gets \Phi^{(t)}(\mathbf{F}_{\mathrm{rx}})$  \Comment{Receive Feature Processor}
    \State $\mathbf{F}_{\mathrm{tx}} \gets \mathbf{H^T} \mathbf{F}_{\mathrm{rx}}$ \Comment{Channel Layer}
    \If{$l > 1$}
        \State$\mathbf{F}_{\mathrm{tx}} = \mathbf{F}_{\mathrm{tx}} + \mathbf{F}_{\mathrm{tx}}^{old}$ \Comment{Skip Connection}
    \EndIf
    \State $\mathbf{F}_{\mathrm{tx}}^{old} \gets \mathbf{F}_{\mathrm{tx}}$
    \State $\mathbf{F}_{\mathrm{tx}} \gets \Psi^{(t)}(\mathbf{F}_{\mathrm{tx}})$  \Comment{Transmit Feature Processor}
    \State $\mathbf{F}_{\mathrm{rx}} \gets \mathbf{H} \mathbf{F}_{\mathrm{tx}}$ \Comment{Channel Layer}
    \State $\mathbf{F}_{\mathrm{rx}} \gets \mathbf{F}_{\mathrm{rx}} - \mathbf{y} \mathbf{1}^T_d$ \Comment{Subtract $\mathbf{y}$}
\EndFor
\State \Return {$\mathbf{F}_{\mathrm{tx}}$}
\EndProcedure
\end{algorithmic}
\end{algorithm}

\textbf{Implementation details:}  Below, we outline two specific designs that enhance the experimental performance of ChannelNet.
\begin{itemize}
    \item \textbf{Skip connection:} Integrating skip connections is a well-established technique to improve gradient flow, thereby mitigating issues such as vanishing or exploding gradients during training \cite{he2016deep}. As shown in Fig. \ref{fig:architecture}a, skip connections are incorporated into the transmit antenna feature processors by adding the input from the $t$-th iteration directly to the input of the $(t+1)$-th iteration.
    
    \item \textbf{Subtraction of $\mathbf{y}$:} The received signal $\mathbf{y}$ is beneficial for guiding the updates of the receive antenna features. While $\mathbf{y}$ is used as the initial receive antenna feature, integrating it into subsequent iterations further enhances performance. An effective method we have identified is to directly subtract $\mathbf{y}$ from the receive antenna feature $\mathbf{F}_{\mathrm{rx}}$ at each feature dimension. Specifically, the input to the receive feature processor is $\mathbf{F}_{\mathrm{rx}} - \mathbf{y} \mathbf{1}^T_d$, where $\mathbf{1}_d$ is an all-ones vector of size $d$.
\end{itemize}

The complete ChannelNet processing procedure is outlined in Algorithm \ref{alg:main}. In contrast to most detectors, ChannelNet offers a significant advantage by eliminating the need for noise power as an input, thereby simplifying the detection process and removing the reliance on accurate noise power estimation.

\textbf{Training process:} 
The training process for ChannelNet involves the optimization of a loss function over a training dataset comprising triples of transmit symbols $\mathbf{x}$, channel matrices $\mathbf{H}$, and received data $\mathbf{y}$. 
This task is formulated as an $M$-class classification problem, where $M$ corresponds to the modulation order used in the system.
To align with the modulation space, the output feature dimension of the transmit feature processor is explicitly set to match the modulation order. Specifically, at the final iteration, the transmit feature processor outputs a matrix \( \mathbf{F}^{(L)}_{\mathrm{tx}} \in \mathbb{R}^{K \times M} \), where each row \( \mathbf{f}_i^{\mathrm{tx}} \) contains unnormalized scores (logits) for the \( M \) modulation symbols at antenna \( i \). These logits are converted into symbol-wise predicted probabilities \( \hat{p}_{i,j} \), which denotes the probability that antenna \( i \) transmitted the \( j \)-th symbol.
The model is trained to minimize the cross-entropy loss between the predicted probabilities and the true labels: 
\begin{equation}
    \mathcal{L} = \sum_{i=1}^{K} \sum_{j=1}^M p_{i, j}\log(\hat{p}_{i, j}),
\end{equation}
where $p_{i, j}$ denotes the ground truth based on the actual transmit symbols $\mathbf{x}$.
Training is carried out using stochastic gradient descent (SGD) to optimize the parameters of $\Phi^{(t)}(\cdot)$ and $\Psi^{(t)}(\cdot)$, minimizing the cross-entropy loss over the training dataset.

\subsection{Invariance and Equivariance to Antenna Permutations} 

ChannelNet-MLP achieves permutation invariance and equivariance through antenna-wise shared feature processors and channel-embedded linear layers. 
Specifically, each antenna feature vector is processed independently through a shared MLP model, ensuring that the processing remains consistent regardless of antenna ordering.
This design preserves the equivariance of antenna features to permutations of the respective antennas. 
The transformations between transmit and receive antenna features are facilitated through channel-embedded linear layers.
Each receive (or transmit) antenna's feature vector is computed as a linear combination of the transmit (or receive) antenna feature vectors, with coefficients from the channel matrix. 
This structure ensures that antenna permutations reorder the coefficient matrix and feature vectors without altering the resulting linear combinations. As a result, the receive (or transmit) antenna feature remains invariant under permutations of the transmit (or receive) antennas. 
By producing transmit antenna features in the final iteration, ChannelNet-MLP ensures that its outputs are equivariant to transmit antenna permutations and invariant to receive antenna permutations.

In contrast, ChannelNet-Conv does not enforce permutation equivariance or invariance due to its convolutional layers. These layers apply filters (kernels) that slide along the antenna dimension, computing dot products with adjacent feature vectors at each position. This sliding mechanism inherently depends on antenna ordering, making the processing sensitive to permutations of the antennas. However, this sensitivity is a strength in scenarios with correlated channels, where spatial dependencies exist between adjacent antennas. In such cases, the convolutional layers effectively capture these local correlations, enabling ChannelNet-Conv to achieve superior performance.

\subsection{Connections with Existing Architectures} 

ChannelNet has connections with both traditional detection algorithms and modern neural architectures, providing effective ways to interpret the model.

\textbf{Connections with iterative detection framework:} ChannelNet shares conceptual similarities with the general iterative detection framework \eqref{equ:iter}.  
According to Algorithm \ref{alg:main}, ChannelNet can be reformulated as an iterative process consisting of two steps\footnote{To highlight the alignment with the general framework \eqref{equ:iter}, this formulation omits the skip connections, which are included to stabilize training. With skip connections, the input to the $\Psi^{(t)}$ in \eqref{equ:iter_ChannelNet} becomes $\mathbf{F}^{(t+\frac{1}{2})}_{\mathrm{tx}} + \mathbf{F}^{(t-\frac{1}{2})}_{\mathrm{tx}}$, rather than $\mathbf{F}^{(t+\frac{1}{2})}_{\mathrm{tx}}$ alone.}:
\begin{equation}\label{equ:iter_ChannelNet}
\begin{aligned}
    \mathbf{F}^{(t+\frac{1}{2})}_{\mathrm{tx}} &= \mathbf{F}^{(t)}_{\mathrm{tx}} + \mathbf{H}^T \Phi^{(t)}\left(\mathbf{H}\mathbf{F}^{(t)}_{\mathrm{tx}} - \mathbf{y} \mathbf{1}^T_d  \right), \\
    \mathbf{F}^{(t+1)}_{\mathrm{tx}} &= \Psi^{(t)} \left(\mathbf{F}^{(t+\frac{1}{2})}_{\mathrm{tx}}\right).
\end{aligned}
\end{equation}
This formulation aligns with \eqref{equ:iter} by setting $\mathbf{A}^{(t)} = \mathbf{H}^T$ and $\mathbf{b}^{(t)} = 0$, while interpreting $\mathbf{F}^{(t+\frac{1}{2})}_{\mathrm{tx}}$ and $\mathbf{F}^{(t)}_{\mathrm{tx}}$ as matrix extensions of the vectors $\mathbf{u}^{(t)}$ and  $\mathbf{x}^{(t)}$, respectively. 
This alignment improves the interpretability of ChannelNet by mirroring the iterative steps of the general framework, providing deeper insights into how information propagates through the model layers and contributes to the final decision-making process.

Although ChannelNet shares some conceptual similarities with this general iterative detection framework  \eqref{equ:iter}, it fundamentally differs in how it enhances detection performance. 
Conventional iterative detectors and model-driven detectors adhering to this framework, such as AMP, OAMP, and MMNet, prioritize the design of sophisticated $\mathbf{A}^{(t)}$ and $\mathbf{b}^{(t)}$ to optimize their detection performance. 
In contrast, ChannelNet improves performance by leveraging powerful data-driven neural networks to learn high-dimensional antenna features and feature processors while utilizing the simplest forms of $\mathbf{A}^{(t)}$ and $\mathbf{b}^{(t)}$. 
High-dimensional feature representations are crucial for neural networks to achieve the expressive capacity required to capture complex channel and signal structures, as supported by theoretical results on the minimum width for universal approximation \cite{lu2017expressive, hanin2017approximating, kidger2020universal}. While AMP extensions such as vector approximate survey propagation (VASP) \cite{chen2023vector} also employ vectorized representations, they follow a statistically derived message-passing framework and differ fundamentally from ChannelNet's end-to-end learning of antenna-specific embeddings.

\textbf{Connections with MLP-Mixer} The proposed ChannelNet-MLP also shares architectural similarities with MLP-Mixer \cite{tolstikhin2021mlp}, a neural network designed for image data processing.  MLP-Mixer alternates between two types of MLPs: one operates on the spatial dimension, mixing information across image patches, while the other operates in the filter dimension, aggregating information within each patch. 
By properly transposing the data and alternating between these two MLP types, MLP-Mixer efficiently captures both spatial and filter-wise dependencies, thereby enhancing the representational capacity of the model for complex image tasks.

ChannelNet-MLP adopts a similar architecture by alternating between two types of MLPs to process features for transmit and receive antennas.
However, ChannelNet-MLP distinguishes itself by incorporating a linear channel layer that transforms the features, enabling effective utilization of the channel matrix.

\section{Expressive Power Analysis}\label{sec:theory}

In this section, we investigate the expressive power of ChannelNet-MLP and demonstrate that it can approximate any continuous permutation-symmetric function and the optimal ML detection function with arbitrary precision under any continuous channel distribution. 

\subsection{Main Results}

We begin by defining the concept of a \textit{permutation-symmetric detection} function as follows:

\begin{definition}[Permutation-Symmetric Detection Function]\label{def:PEF}
    A function $f: \mathbb{R}^{N \times K} \times \mathbb{R}^N  \rightarrow \mathbb{R}^K, (\mathbf{H}, \mathbf{y}) \rightarrow \mathbf{\hat{x}}$ is said to be permutation-symmetric if, for any permutations $\sigma_{\mathrm{rx}} \in \mathcal{S}_N$ and $\sigma_{\mathrm{tx}} \in \mathcal{S}_K$, it satisfies
    \begin{equation}
        \sigma_{\mathrm{tx}}\left(f(\mathbf{H}, \mathbf{y})\right) = f\left(\mathbf{H}_{\sigma_{\mathrm{rx}}, \sigma_{\mathrm{tx}}}, \mathbf{y}_{\sigma_{\mathrm{rx}}}\right),
    \end{equation}  
where $\sigma_{\mathrm{tx}}\left(f(\mathbf{H}, \mathbf{y})\right)$ denotes the vector obtained by permuting the elements of $f(\mathbf{H}, \mathbf{y})$ according to $\sigma_{\mathrm{tx}}$,  $\mathbf{H}_{\sigma_{\mathrm{rx}}, \sigma_{\mathrm{tx}}}$ represents the matrix $\mathbf{H}$ with its rows permuted by $\sigma_{\mathrm{rx}}$ and its columns permuted by $\sigma_{\mathrm{tx}}$, and $\mathbf{y}_{\sigma_{\mathrm{rx}}}$ represents the vector $\mathbf{y}$ with its entries permuted by $\sigma_{\mathrm{rx}}$.
\end{definition}

This definition characterizes permutation-symmetric detection functions as those that produce equivariant outputs with respect to permutations of the transmit antennas and invariant outputs with respect to permutations of the receive antennas. In the following theorem, we demonstrate that ChannelNet-MLP can effectively approximate any continuous permutation-symmetric detection function under any continuous channel distribution.

\textcolor{black}{\begin{thm}\label{thm:main}
Let $p(\mathbf{H, y})$ be a continuous probability density function over $\mathbb{R}^{N \times K} \times \mathbb{R}^N $ and let $g$ be any continuous permutation-symmetric detection function.  For any given $\epsilon > 0$ and $0<\delta<1$, there exists a function $f$ represented by ChannelNet-MLP, such that the probability that their output difference is less than $\epsilon$ exceeds $1-\delta$. Specifically,  for $(\mathbf{H, y}) \sim p(\mathbf{H, y})$,
\begin{equation*}
    \mathbb{P}(\|g(\mathbf{H, y}) - f(\mathbf{H, y})\| < \epsilon) > 1- \delta.
\end{equation*}
\end{thm}
All proofs are deferred to Appendix A.}

This theorem highlights the richness of the function space represented by ChannelNet-MLP, making it capable of approximating any continuous permutation-symmetric detection functions.
In the context of MIMO detection, the ML detector represents the optimal detection method. 
In the following corollary, we establish that ChannelNet-MLP can approximate the ML detector, even though the ML detection function is inherently  discrete rather than continuous.

\textcolor{black}{\begin{col}
Let $p(\mathbf{H, y})$ be a continuous probability density function over $\mathbb{R}^{N \times K} \times \mathbb{R}^N $. For any given $\epsilon > 0$ and $0<\delta<1$, there exists a function $f$ represented by ChannelNet-MLP, such that, for $(\mathbf{H, y}) \sim p(\mathbf{H, y})$, 
\begin{equation*}
    \mathbb{P}(\|f_{ML}(\mathbf{H, y}) - f(\mathbf{H, y})\| < \epsilon) > 1 - \delta,
\end{equation*}
\end{col}
where $f_{ML}$ denotes the ML detector.}

The universal approximation property of ChannelNet-MLP builds on the foundational result that MLPs can approximate any continuous function with arbitrary precision, given sufficient model complexity. This property highlights ChannelNet-MLP's potential to match the performance of the ML detector under any channel distribution, offering a flexible and powerful framework for MIMO detection. However, realizing this potential in practice often demands substantial model complexity and extensive training data, which may not always be feasible. Obtaining theoretical performance guarantees with a limited model size and fewer training samples remains an open challenge.

\section{Experiments}\label{sec:experiment}
In this section, we present experimental results comparing the performance of ChannelNet with other state-of-the-art massive MIMO detectors, using symbol error rate (SER) as the evaluation metric. We assess SER under various conditions, including different numbers of antennas, modulation orders, and channel distributions. Below are the implementation details of the detection schemes used in our experiments:

\begin{itemize}
    \item \textbf{MMSE:} A classical linear detector that applies the channel-noise regularized pseudo-inverse of the channel matrix to invert the signal and rounds the output to the nearest constellation point.
    \item \textbf{AMP:} An iterative detection algorithm \cite{jeon2015optimality}, implemented with 50 iterations, as the performance saturates beyond this point according to \cite{khani2020adaptive}.
    \item \textbf{V-BLAST:} A multi-stage successive interference cancellation algorithm using the ZF detector at the detection stage, as presented in \cite{wolniansky1998v}.
    \item \textbf{ML:} The optimal detector, implemented using the Gurobi optimization package \cite{optimization2020gurobi}.
    \item \textbf{DetNet\footnote{We use the official repository at \url{https://github.com/neevsamuel/DeepMIMODetection/tree/master}}:} A learning-based detector with  $3N_t$ layers, as described in \cite{samuel2017deep}.
    \item \textbf{OAMPNet\footnote{We use the official repository at \url{https://github.com/hehengtao/OAMP-Net}}:} A learning-based detector that unfolds OAMP, with two learnable parameters per iteration \cite{he2020model}. 
    \item \textbf{AMP-GNN\footnote{We use the official repository at \url{https://github.com/hehengtao/AMP_GNN}}:} A learning-based detector that introduces a GNN module in unfolding AMP \cite{he2023message}. 
    \item \textbf{RE-MIMO\footnote{We use the official repository at \url{https://github.com/krpratik/RE-MIMO}}:} A learning-based detector employing an encoder-predictor architecture, with the encoder parameterized by transformers and the predictor by MLPs \cite{pratik2020re}. 
    \item \textbf{ChannelNet-MLP:}  The proposed detector as explained in Section \ref{sec:methodology}. It is implemented with $20$ iterations and a feature dimension of $d = 10$. The detailed network architecture is provided in Appendix~C. The model is trained for $600$ epochs with $2 \times 10^6$ samples per epoch, using a batch size of $64$.   The Adam optimizer \cite{kingma2014adam} with a momentum of $0.9$ is used, starting with an initial learning rate of $10^{-3}$, which is reduced by a factor of $0.1$ every $200$ epochs.  
    \item \textbf{ChannelNet-Conv:} Two additional convolutional layers are added to each antenna feature processor of ChannelNet-MLP.  The complete architecture is also described in Appendix~C.
\end{itemize}

For all the experiments in this section, we consider two modulation schemes, namely 16-QAM and 64-QAM, along with two antenna configurations, $(N_r, N_t) = (64, 32)$ and $(N_r, N_t) = (128, 64)$. The performance is evaluated across three types of channels: Rayleigh fading channels, correlated channels, and 3GPP channels. Unless otherwise specified, the channel noise follows an \textit{i.i.d.} zero-mean Gaussian distribution, and the noise power is determined by the SNR according to the formula:
\begin{equation}
    \text{SNR} = \frac{\mathbb{E}[\|\mathbf{Hx}\|^2_2]}{\mathbb{E}[\|\mathbf{n}\|^2_2]}. 
\end{equation}
Of all the detectors evaluated, only ChannelNet and ML operate without requiring noise power as input. In contrast, the other detectors in our experiments are assessed under the assumption of precisely known noise power.

\begin{figure*}[t] 
\centering
\begin{subfigure}{0.49\textwidth}
    \includegraphics[width=\textwidth]{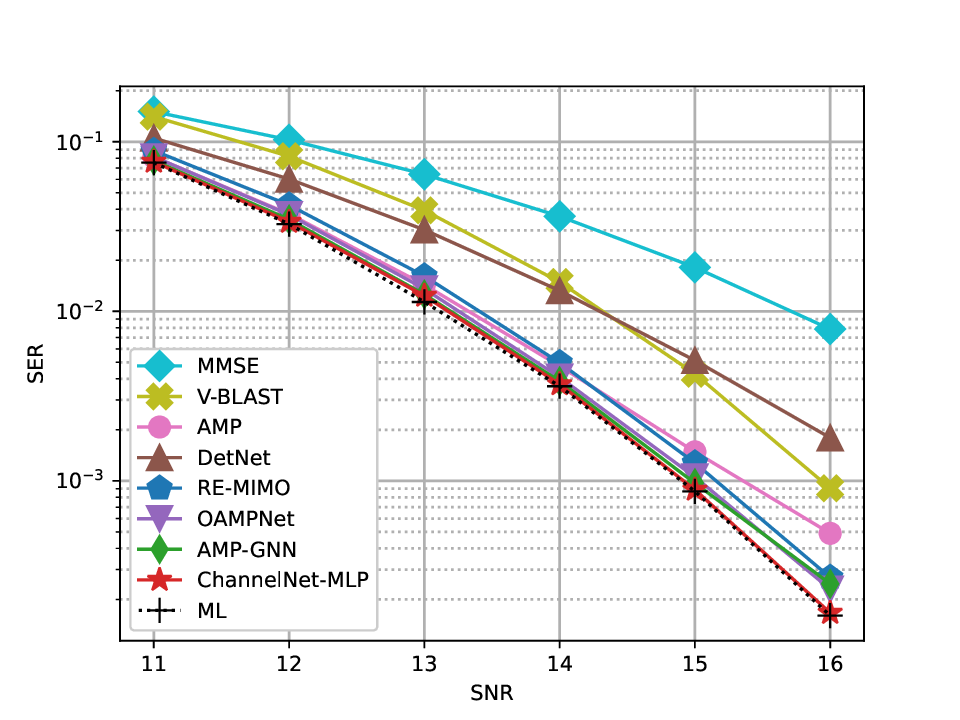}
    \label{fig:first}
\end{subfigure}
\begin{subfigure}{0.49\textwidth}
    \includegraphics[width=\textwidth]{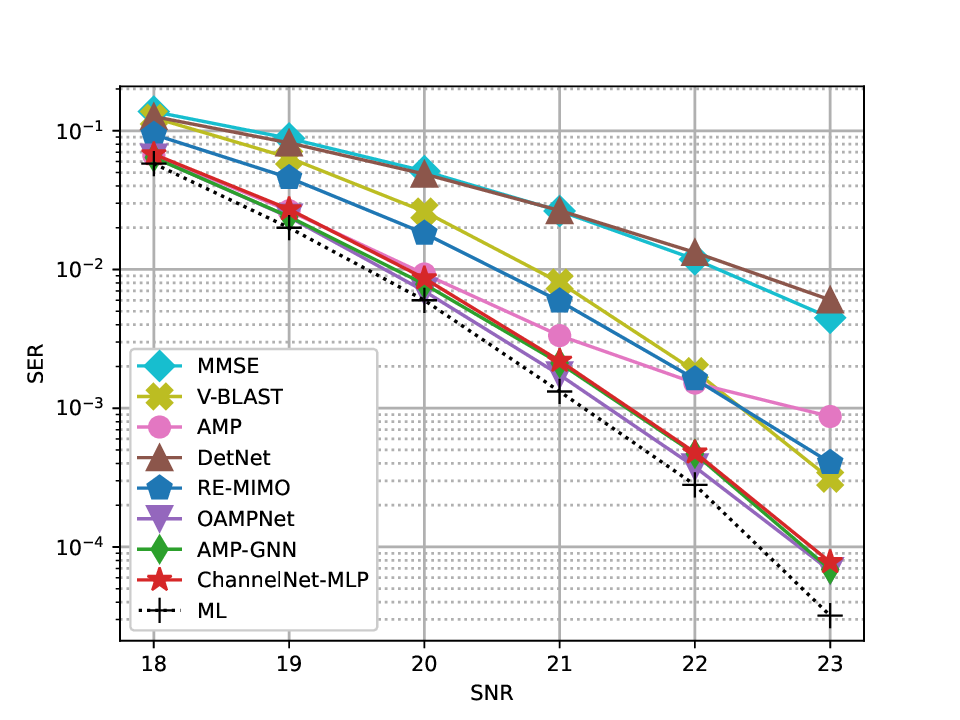}
    \label{fig:second}
\end{subfigure}
\caption{SER performance under Rayleigh fading channels for $(N_r, N_t) = (64, 32)$ with 16-QAM (left) and 64-QAM (right).}\label{fig:iid_32x64}
\end{figure*}

\begin{figure*}[h] 
\centering
\begin{subfigure}{0.49\textwidth}
    \includegraphics[width=\textwidth]{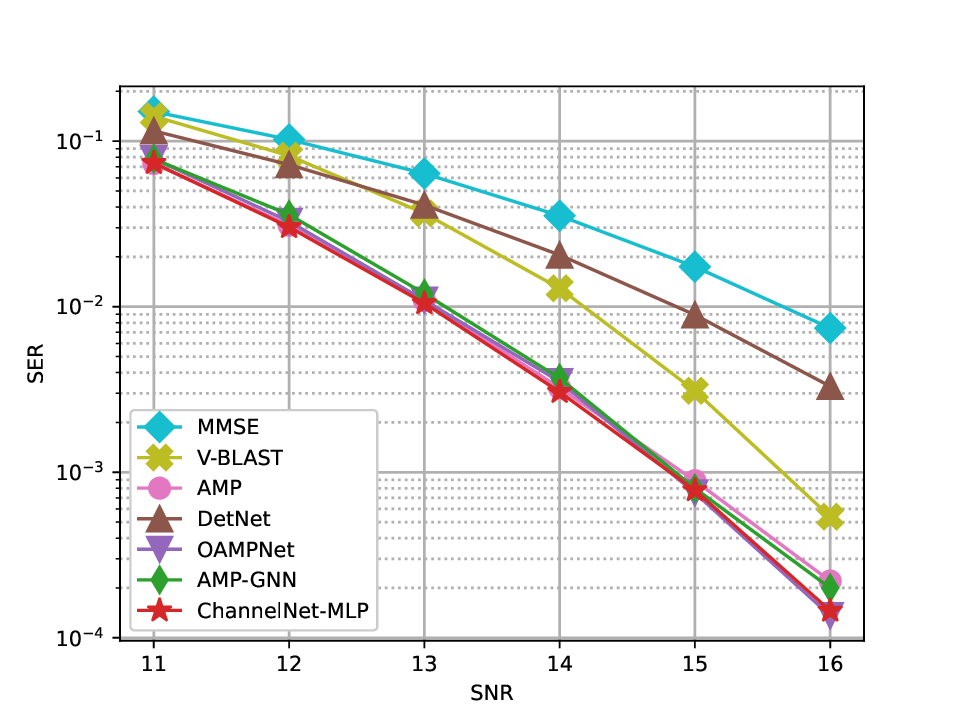}
    \label{fig:first}
\end{subfigure}
\begin{subfigure}{0.49\textwidth}
    \includegraphics[width=\textwidth]{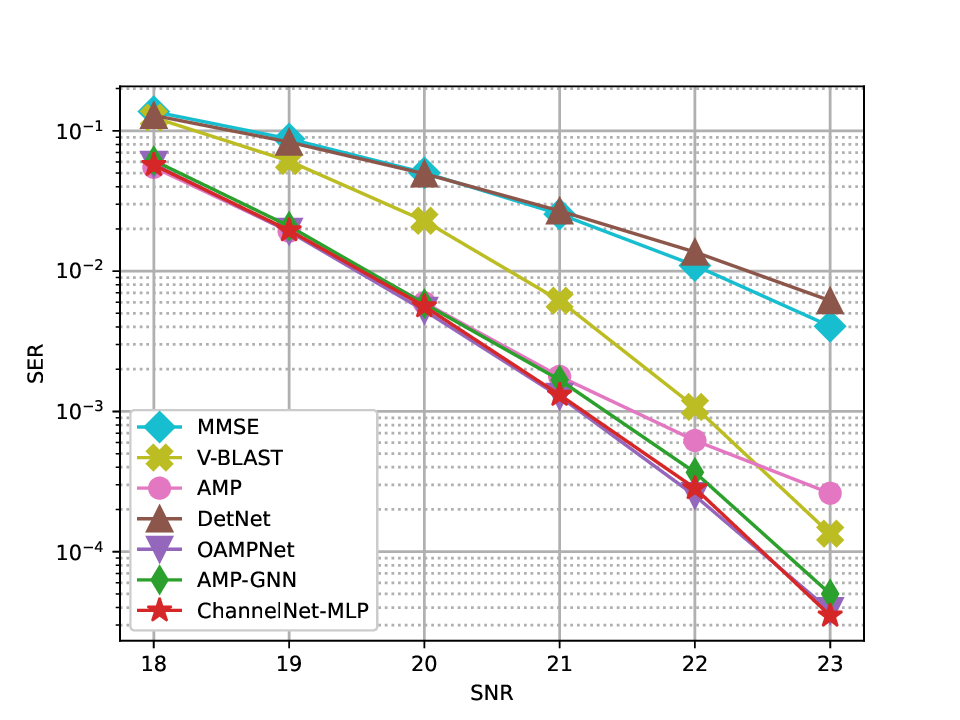}
    \label{fig:second}
\end{subfigure}
\caption{SER performance under Rayleigh fading channels for $(N_r, N_t) = (128, 64)$ with 16-QAM (left) and 64-QAM (right).}\label{fig:iid_64x128}
\end{figure*}

\subsection{Performance under Various Channels}
We evaluate ChannelNet's performance across Rayleigh fading, correlated, and 3GPP channels. In each experiment, the training and testing samples are drawn from the same channel distribution. 

\subsubsection{Rayleigh Fading Channels}
We first consider a Rayleigh fading channel, where each element in $\mathbf{\tilde{H}}$ follows a zero-mean circularly-symmetric Gaussian distribution with variance $1/N_r$, \textit{i.e}., $\mathbf{\tilde{H}}_{ij} \sim \mathcal{CN}(0, 1/N_r)$. This assumption aligns with the characteristics of an \textit{i.i.d.} Gaussian channel. Due to the \textit{i.i.d.}. nature of the channel, the permutation of antennas does not affect the channel distribution. Therefore, only ChannelNet-MLP is evaluated for comparison.

Fig. \ref{fig:iid_32x64} presents a comprehensive comparison of the SER for ChannelNet-MLP and other detectors under 16-QAM and 64-QAM modulations with an antenna configuration of $(N_r, N_t) = (64, 32)$. ChannelNet-MLP consistently matches or outperforms other detectors, and its performance is close to the ML bound, showcasing its capability in approximating optimal performance. In contrast, traditional detectors, such as MMSE and V-BLAST, exhibit significantly higher SER, especially at higher SNRs. Although DetNet exhibits improved performance over traditional methods, it still falls short of ChannelNet-MLP. RE-MIMO achieves performance comparable to ChannelNet-MLP with 16-QAM but falls behind under 64-QAM. OAMPNet and AMP-GNN exhibit similar performance to ChannelNet-MLP in both modulation schemes, with all methods achieving near-ML results.

Fig. \ref{fig:iid_64x128} illustrates the SER comparison for $(N_r, N_t) = (128, 64)$. Results for RE-MIMO and ML are omitted due to scalability limitations.  The performance trends observed in this larger antenna setting are consistent with those in the $(N_r, N_t) = (64, 32)$ case,  where  OAMPNet, AMP-GNN, and ChannelNet-MLP achieve the best performance compared with other detectors. This consistency underscores the scalability of ChannelNet-MLP across varying numbers of antennas.

\begin{figure*}[t] 
\centering
\begin{subfigure}{0.49\textwidth}
    \includegraphics[width=\textwidth]{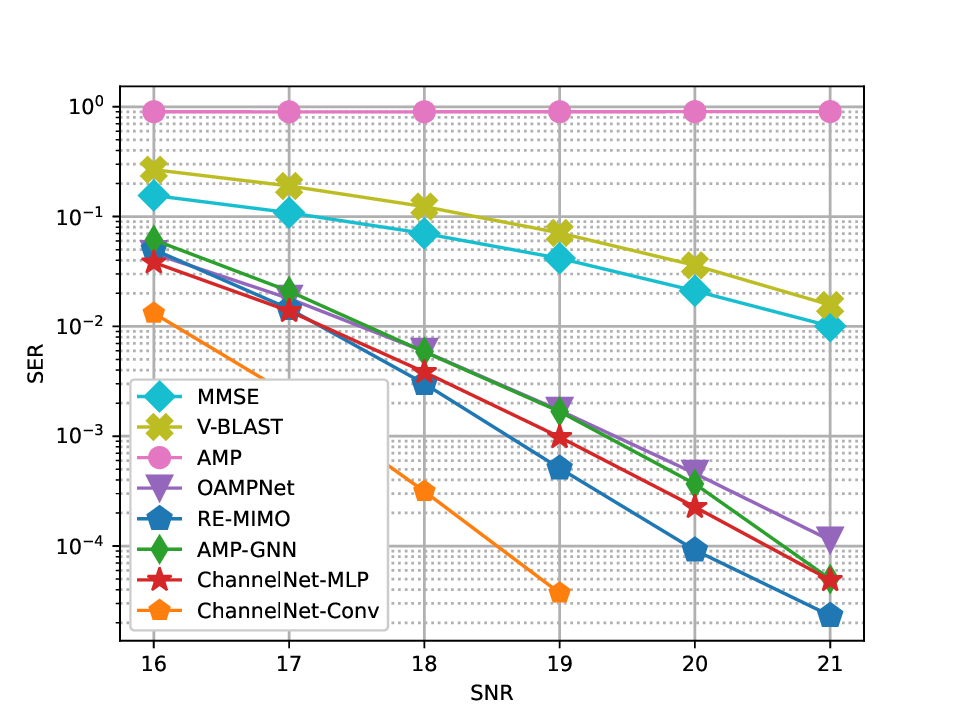}
    \label{fig:first}
\end{subfigure}
\begin{subfigure}{0.49\textwidth}
    \includegraphics[width=\textwidth]{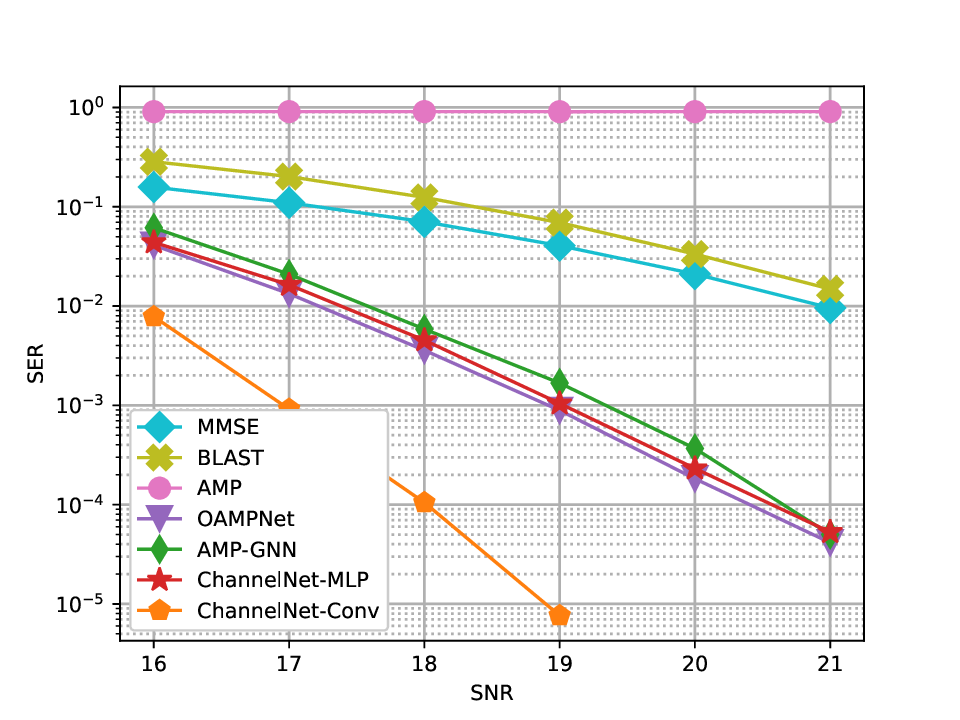}
    \label{fig:second}
\end{subfigure}
\caption{SER performance under correlated channels for $(N_r, N_t) = (64, 32)$ (left) and $(N_r, N_t) = (128, 64)$ (right) with 16-QAM.}\label{fig:cor_32x64}
\end{figure*}

\subsubsection{Correlated Channels}
In this experiment, we evaluate ChannelNet's performance under correlated MIMO channels, modeled using the Kronecker model:
\begin{equation}
    \mathbf{H} = \mathbf{R_R}^{1/2}\mathbf{H_w}\mathbf{R_T}^{1/2},
\end{equation}
where $\mathbf{H_w}$ represents the Rayleigh fading channel matrix, and $\mathbf{R_R}$ and $\mathbf{R_T}$ denote the spatial correlation matrices at the receiver and transmitter, respectively. These matrices are generated according to the exponential correlation model \cite{loyka2001channel}, with a fixed correlation coefficient of $0.6$. Due to the spatial correlations among antennas, the permutation of antennas affects the channel distribution. Accordingly, we evaluate both ChannelNet-MLP and ChannelNet-Conv to assess their effectiveness under these conditions.

Fig. \ref{fig:cor_32x64} illustrates the SER performance for two antenna configurations, $(N_r, N_t) = (64, 32)$ and $(N_r, N_t) = (128, 64)$,  under 16-QAM modulation.  Conventional detectors, including MMSE, V-BLAST, and AMP, perform significantly worse than the learning-based methods.  Among the learning-based detectors, ChannelNet-MLP achieves performance comparable to the model-driven baselines, while  ChannelNet-Conv surpasses both ChannelNet-MLP and the model-driven baselines by 1-2 dB. 
This performance gain can be attributed to ChannelNet-Conv's convolutional layers, which effectively capture spatial correlations, thereby improving detection accuracy.

\subsubsection{3GPP Channels} 

Finally, we examine the performance of ChannelNet under the 3GPP 3D MIMO channel model \cite{3gpp} simulated using the QuaDRiGa channel simulator \cite{jaeckel2014quadriga}. Specifically, the simulation is conducted under the ``3D-UMa-NLOS'' scenario, representative of urban environments with obstructed line-of-sight due to buildings. In this setup, a base station with a $64$-element single-polarized antenna array is positioned $20$ meters above ground and covers a sector with a $500$-meter radius. Channel samples are collected by randomly distributing 32 single-polarized omni-directional antennas within the coverage area. The simulation operates at a carrier frequency of $3.5$ GHz. 

\begin{figure*}[t] 
\centering
\begin{subfigure}{0.49\textwidth}
    \includegraphics[width=\textwidth]{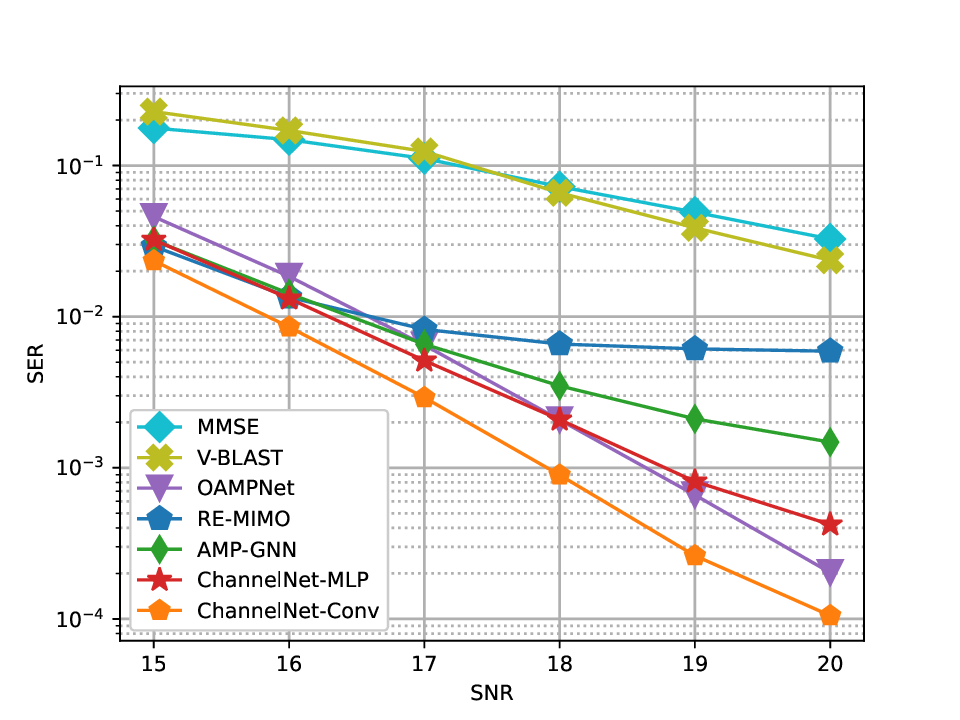}
    \subcaption{}
    \label{fig:result_3gpp_a}
\end{subfigure}
\begin{subfigure}{0.49\textwidth}
    \includegraphics[width=\textwidth]{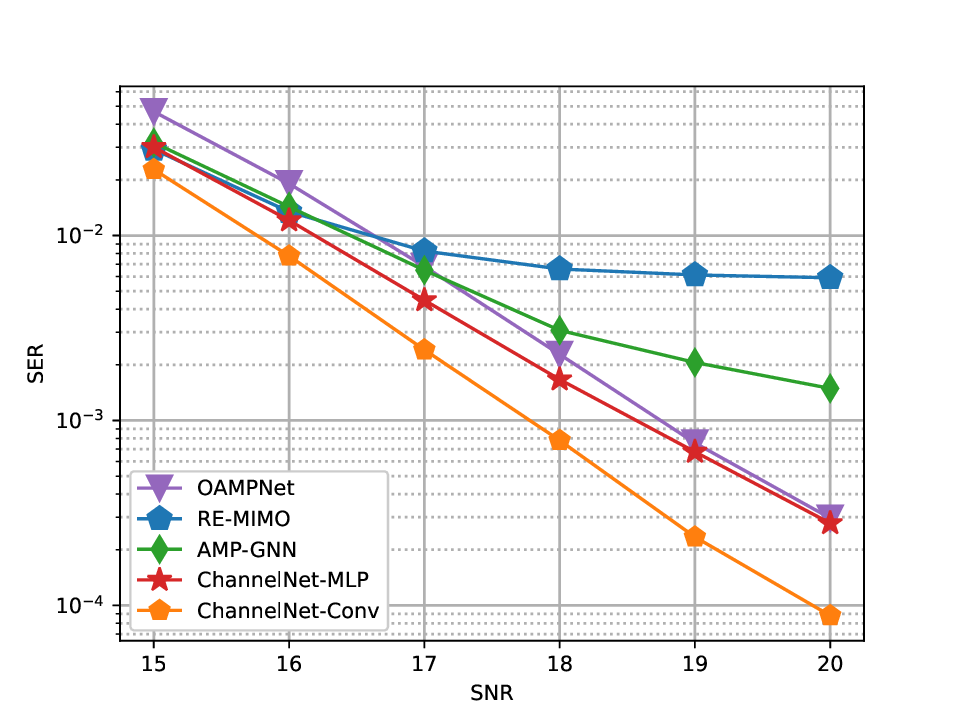}
    \subcaption{}
    \label{fig:result_3gpp_b}
\end{subfigure}
\caption{SER performance under 3GPP channels for $(N_r, N_t) = (64, 32)$ with 16-QAM. Models are trained on ``3GPP-UMa-NLOS'' channels and tested with in-distribution channels (left, ``3D-UMa-NLOS'') and out-of-distribution channels (right, ``3D-UMi-NLOS'').}\label{fig:result_3gpp}
\end{figure*}

Fig. \ref{fig:result_3gpp_a} presents the SER performance under 16-QAM modulation. Similar to the correlated channel scenario, conventional detectors perform significantly worse than learning-based methods. Among the learning-based approaches, ChannelNet-MLP achieves performance on par with OAMPNet, outperforming the other methods, while ChannelNet-Conv exhibits the best overall performance, outperforming both OAMPNet and ChannelNet-MLP by approximately 1 dB.

\subsection{Robustness Evaluation}

We evaluate the robustness of ChannelNet by testing it in conditions that differ from the training environment. We consider scenarios with channel distribution shifts, channel estimation errors, and non-Gaussian noise. In each case, the model is trained under ideal conditions and then evaluated under these adverse scenarios to determine its robustness.

\begin{figure*}[t] 
\centering
\begin{subfigure}{0.49\textwidth}
    \includegraphics[width=\textwidth]{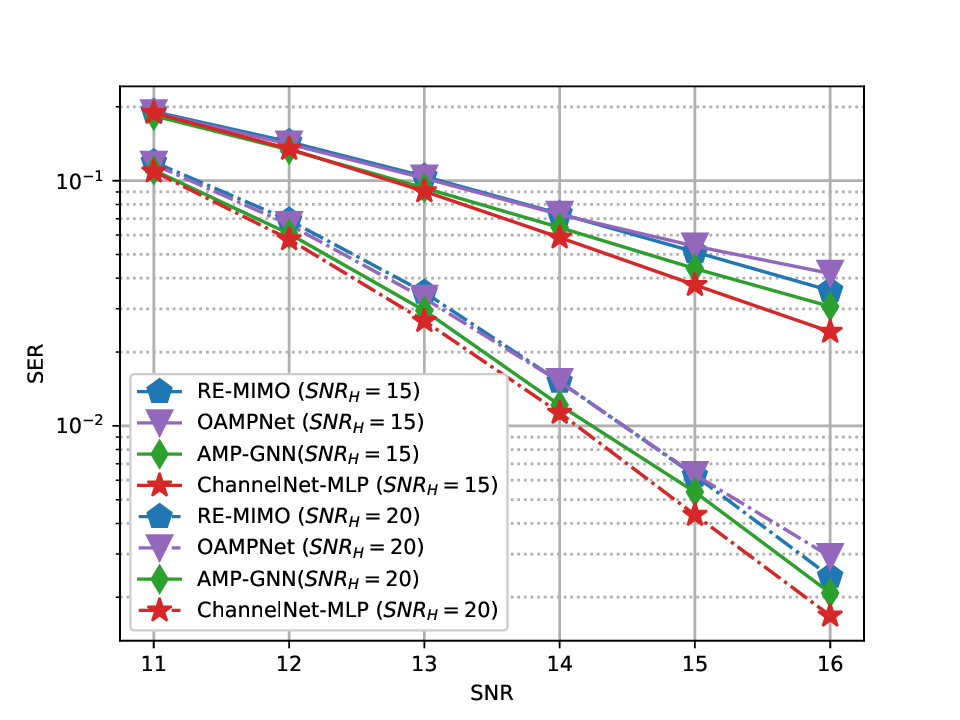}
    \subcaption{}
    \label{fig:robust_channel_error}
\end{subfigure}
\begin{subfigure}{0.49\textwidth}
    \includegraphics[width=\textwidth]{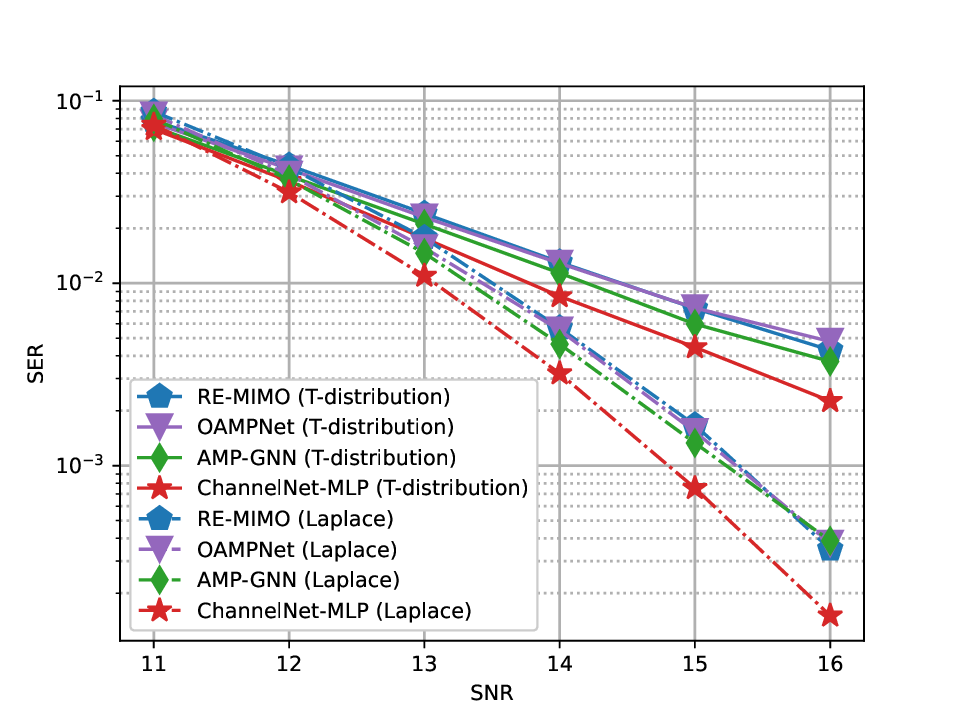}
    \subcaption{}
    \label{fig:robust_noise}
\end{subfigure}
\caption{SER performance of models trained on Rayleigh fading channels with $(N_r, N_t) = (64, 32)$ and 16-QAM modulation, evaluated under channel estimation errors (left) and non-Gaussian noise scenarios (right). }\label{fig:robust_32x64}
\end{figure*}

\subsubsection{Channel Distribution Shift}
We first assess the robustness of the detectors under channel distribution shifts, a critical factor in real-world communication systems where channel distributions often deviate from training environments. 
The 3GPP channel model offers a range of scenarios tailored for different communication environments, making it well-suited for this evaluation. Specifically, we train all detectors using channels from the ``3D-UMa-NLOS'' scenario, but test them with channels from the ``3D-UMi-NLOS'' scenario to simulate a channel distribution shift.  Fig. \ref{fig:result_3gpp_b} illustrates the SER performance of ChannelNet compared to model-driven detectors under the channel distribution shift. By comparing Fig. \ref{fig:result_3gpp_a} and Fig. \ref{fig:result_3gpp_b}, it is evident that ChannelNet exhibits superior robustness to channel distribution shifts. Under the shifted conditions, ChannelNet-MLP outperforms all model-driven detectors, whereas in Fig. \ref{fig:result_3gpp_a}, its performance is comparable to that of OAMPNet. Additionally, the performance gap between ChannelNet-Conv and the model-driven detectors becomes more pronounced under the shifted conditions, further highlighting ChannelNet's robustness.

\subsubsection{Channel Estimation Noise}
In practical systems, perfect channel state information (CSI) is often unattainable. To account for this limitation,  we add noise to the channel matrix $\mathbf{H}$, resulting in a perturbed matrix:
\begin{equation}
    \hat{\mathbf{H}} = \mathbf{H} + \mathbf{E},
\end{equation}
where $\hat{\mathbf{H}}$ represents the perturbed channel matrix, $\mathbf{H}$ the true channel matrix, and $\mathbf{E}$ the channel estimation error. Each element of $\mathbf{E}$ is drawn from a zero-mean \textit{i.i.d.} Gaussian distribution, with its variance determined by the SNR of CSI, defined as:
\begin{equation}
   \text{SNR}_{\mathbf{H}} = \frac{\|\mathbf{H}\|^2_F}{\|\mathbf{E}\|^2_F},
\end{equation}
where $\|\cdot\|_F$ denotes the Frobenius norm.

To evaluate the robustness of the detectors to channel estimation errors, all models are trained without such errors and tested under conditions where errors are introduced. Specifically, the training is conducted on Rayleigh fading channels with an antenna configuration of $(N_r, N_t) = (64, 32)$ using 16-QAM modulation. Fig. \ref{fig:robust_channel_error} presents the SER performance of ChannelNet-MLP, OAMPNet, AMP-GNN, and RE-MIMO under channel estimation error scenarios. Two levels of $\text{SNR}_{\mathbf{H}}$ are considered: $15$ dB and $20$ dB. The results show that ChannelNet-MLP achieves superior performance in both cases. Given that ChannelNet-MLP demonstrates performance comparable to the other detectors with consistent training and test conditions as shown in Fig. \ref{fig:iid_32x64}, the observed differences in Fig. \ref{fig:robust_channel_error} highlight that ChannelNet-MLP offers enhanced robustness compared to the model-driven baselines.

\subsubsection{Non-Gaussian Channel Noise}

While Gaussian noise models are suitable for various practical scenarios, they do not capture the characteristics of disturbances prevalent in environments such as urban, indoor, and underwater settings, where non-Gaussian noise is common.

To evaluate robustness to non-Gaussian noise, detectors are trained with Gaussian noise but tested under non-Gaussian noise conditions. We continue to use models that are trained on Rayleigh fading channels with an antenna configuration of $(N_r, N_t) = (64, 32)$ and 16-QAM modulation. Two non-Gaussian noise distributions are considered during evaluation: the $t$-distribution with $\nu=3$ and the Laplace distribution. The SNR is calculated as the ratio of signal to noise power, where the non-Gaussian noise power is derived from the variance of the corresponding distribution. Fig. \ref{fig:robust_noise} illustrates the SER performance of ChannelNet-MLP, OAMPNet, AMP-GNN, and RE-MIMO under non-Gaussian noise conditions. ChannelNet-MLP outperforms the other detectors in these scenarios.  Given its comparable performance to other detectors under Gaussian noise, this improvement under non-Gaussian noise underscores its enhanced robustness to noise distributions.

\subsection{Impact of Feature Dimension and Iteration Number}
ChannelNet is a purely data-driven method with standard deep learning components, allowing for easy adjustments to the model size to achieve a tradeoff between model complexity and performance. In this subsection, we demonstrate the impact of the number of iterations $L$ and the feature dimension $d$. Our evaluations are conducted using Rayleigh fading channel with $(N_r, N_t) = (64, 32)$ and 64-QAM modulation.

\begin{figure}[t!]
    \centering
    \includegraphics[width=1\linewidth]{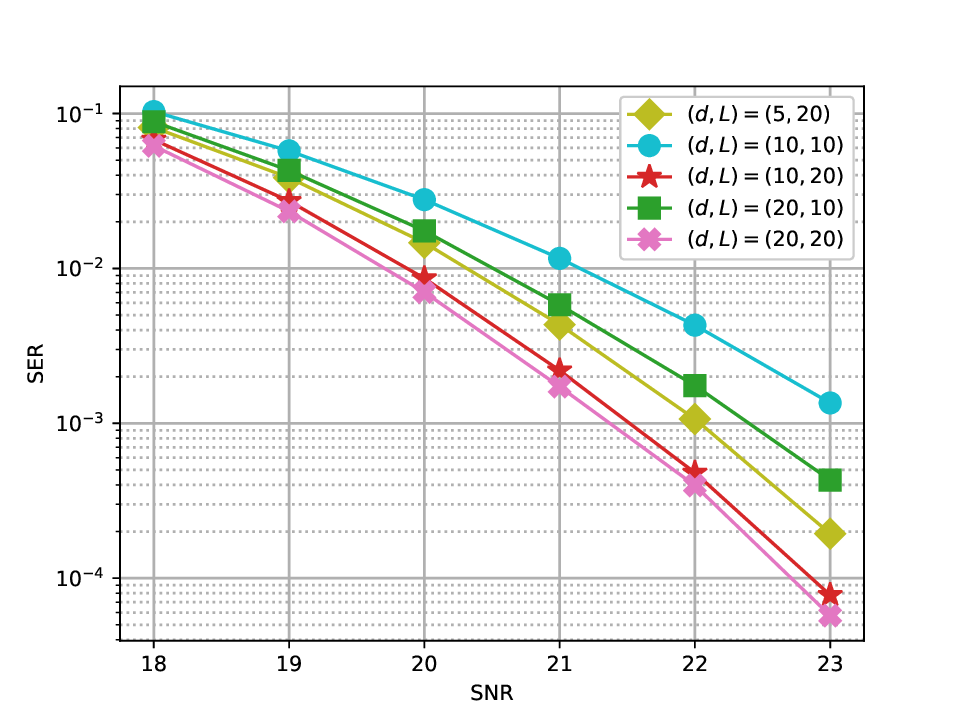}
    \caption{SER performance for various feature dimensions $d$ and iterations $L$ under  Rayleigh fading channel with $(N_r, N_t) = (64, 32)$ and 64-QAM modulation.}
    \label{fig:tradeoff}
\end{figure}

Fig.~\ref{fig:tradeoff} presents the performance of ChannelNet under various configurations. When comparing to the baseline model with $(d,L)=(10,20)$, increasing the feature dimension to $d=20$ results in a slight performance improvement. In contrast, decreasing either the feature dimension or the number of iterations results in a significant performance degradation, highlighting the importance of both sufficiently rich feature representations and adequate iterative refinement. Moreover, the comparison between $(d, L) = (10, 20)$ and $(20, 10)$ indicates that performance is more sensitive to the number of iterations than to the feature dimension.


\begin{table*}[t!]
    \centering
    \resizebox{\textwidth}{!}{
    \begin{tabular}{c|c|c|c|c|c|c|c|c|c|c} \hline \hline
        \textbf{Detector} & \textbf{Asymptotic Complexity} & \multicolumn{3}{c|}{\textbf{Multiplications} $(\times 10^7)$ } & \multicolumn{3}{c|}{\textbf{Time (ms)} (Inference/train per batch)} \\ \cline{3-8}
        & & \makecell[c]{\textbf{64-QAM} \\ $N_t=32$, \\ $N_r=64$} & \makecell[c]{\textbf{64-QAM} \\ $N_t=64$, \\ $N_r=128$} & \makecell[c]{\textbf{64-QAM} \\ $N_t=128$, \\ $N_r=256$} & \makecell[c]{\textbf{64-QAM} \\ $N_t=32$, \\ $N_r=64$} & \makecell[c]{\textbf{64-QAM}\\ $N_t=64$, \\ $N_r=128$} & \makecell[c]{\textbf{64-QAM} \\ $N_t=128$, \\ $N_r=256$} \\
        \hline
        AMP & $\mathcal{O}(N_t N_r) $ & 0.18  &  0.67 & 2.68 & 1.64 & 1.71 & 2.11 \\

        MMSE & $\mathcal{O}(N^3_t + N_t^2N_r)$ & 0.080 & 0.63 &  5.05 & 0.18 & 0.26 & 0.64 \\
        AMP-GNN &  $\mathcal{O}(N_t N_r + N_t^2)$ & 3.98 & 15.90 & 63.53 & 12.4/33.1  & 94.8/335.4 & 1320.1/2994.4\\
        OAMPNet & $\mathcal{O}(N^3_t + N^3_r + N_t^2N_r + N_r^2N_t)$ & 1.91  &  15.2 & 121.1 & 3.88/4.76 & 13.3/14.2 & 76.8/78.3 \\
        RE-MIMO & $\mathcal{O}(N_t^2N_r + N_r^2N_t)$  & 59.1  & 159.7 & 517.5 & 4.42/9.98 & 12.1/25.68 & 184.5/406.58\\
        
        \hline
        \textbf{ChannelNet-MLP} & $\mathcal{O}(N_t N_r)$ & 0.98 & 1.22 & 3.48 & 0.88/2.35 & 0.94/2.45 & 1.07/2.79 \\
        \textbf{ChannelNet-Conv} & $\mathcal{O}(N_t N_r)$ & 1.87  & 2.20 & 4.51 & 1.96/5.24 & 2.20/5.86 & 2.56/6.30 \\
        \hline
    \end{tabular}}
    \caption{Computational complexity comparison of massive MIMO detectors.}
    \label{tab:complexity}
\end{table*}
\subsection{Complexity Analysis} 
\textbf{Number of Multiplications:} 
The computational complexity of ChannelNet is primarily determined by its channel layers and antenna feature processors. 
Each channel layer involves a matrix multiplication, either between a $K \times N$ matrix and an $N \times d$ matrix or between an $N \times K$ matrix and a $K \times d$ matrix, with both cases requiring $d N K$ multiplications. 
In ChannelNet-MLP, the antenna feature processors are implemented using MLPs with one hidden layer containing $d$ neurons. 
Each MLP requires $2d^2$ multiplications, resulting in a total of $2d^2K$ multiplications for the transmit antenna processor and $2d^2N$ multiplications for the receive antenna processor per iteration.  Therefore, the overall number of multiplications for ChannelNet-MLP is $L(dNK + 2d^2K + 2d^2N)$, where $L$ represents the number of iterations.
For ChannelNet-Conv, additional complexity arises from the two convolutional layers added before the MLPs.  These layers require $2kd^2K$ and $2kd^2N$ multiplications for each transmit and receive processor, respectively, where $k$ is the  kernel size. Consequently, the total number of multiplications for ChannelNet-Conv is $L(d N K + (2+2k)d^2 K + (2+2k) d^2 N)$. 
Given that both $d$, $k$, and $L$ are constants independent of $N$ and $K$, the asymptotic computational complexity for both ChannelNet-MLP and ChannelNet-Conv can be expressed as $\mathcal{O}(N K)$, or equivalently, $\mathcal{O}(N_t N_r)$.

Table \ref{tab:complexity} compares the asymptotic computational complexities of the detectors. From the table, both AMP and ChannelNet exhibit the lowest asymptotic computational complexity, $\mathcal{O}(N_t N_r)$. In contrast, the MMSE detector demands a complexity of $\mathcal{O}(N^3_t + N_t^2N_r)$ due to matrix pseudo-inversion, making it less suitable for high-dimensional scenarios. However, it remains efficient for small to moderate numbers of antennas because of its non-iterative nature.
Model-driven detectors, including AMP-GNN, OAMPNet, and RE-MIMO, exhibit higher asymptotic computational complexities compared to AMP and ChannelNet.
Specifically, AMP-GNN's complexity of $\mathcal{O}(N_t N_r + N_t^2)$ arises from the AMP iterations and the GNN component.  OAMPNet has a complexity of $\mathcal{O}(N^3_t + N^3_r + N_t^2N_r + N_r^2N_t)$ due to matrix pseudo-inversion in each iteration, while RE-MIMO's complexity of $\mathcal{O}(N_t^2 N_r + N_r^2 N_t)$ reflects the computational demands of its attention mechanisms.

In addition to analyzing the asymptotic complexity, we compare the exact number of multiplications required for specific configurations, as summarized in Table \ref{tab:complexity}. ChannelNet-Conv requires approximately twice the number of multiplications as ChannelNet-MLP. Compared to traditional detectors such as AMP and MMSE, ChannelNet's computational demands become increasingly competitive as the number of antennas grows. This efficiency is particularly evident when compared to learning-based model-driven detectors. For instance, with $(N_r, N_t) = (256, 128)$, ChannelNet-Conv requires approximately $14$, $27$, and $114$ times fewer multiplications than AMP-GNN, OAMPNet, and RE-MIMO, respectively. 

\textbf{Inference and Training Time:} We also compare the inference and training times of the detectors, with all algorithms implemented in PyTorch and tested on an NVIDIA 4090 GPU. The times are measured for a batch of $256$ samples. As expected, ChannelNet-Conv takes approximately twice as long as ChannelNet-MLP due to its additional convolutional layers. The inference times for both ChannelNet-MLP and ChannelNet-Conv are comparable to those of traditional detectors, especially in larger systems, mirroring their comparability in the number of multiplications. For learning-based methods, training times are typically about twice as long as the inference times due to the forward and backward passes involved.  
Notably, both ChannelNet-MLP and ChannelNet-Conv demonstrate significantly lower inference and training times compared to other learning-based detectors, highlighting their computational efficiency and suitability for large-scale massive MIMO scenarios.

\textbf{Extension to Wideband Channels:} While this work focuses on narrowband channels, ChannelNet can be readily extended to wideband systems by incorporating orthogonal frequency division multiplexing (OFDM) and applying the detector independently across subcarriers, which results in a linear increase in total computation. However, this increase can be effectively mitigated through parallel processing, where subcarriers can be processed concurrently. Furthermore, the training procedure naturally accommodates this extension via batched processing without introducing additional overhead.

\section{Conclusion and Future Directions}\label{sec:conclusion}
In this paper, we present ChannelNet, a purely data-driven learning-based massive MIMO detector.
In contrast to model-driven learning-based detectors,  ChannelNet breaks free from conventional iterative frameworks and assumptions regarding  channel and noise distribution.
Our theoretical analysis demonstrates that it can approximate the optimal ML detector with arbitrary precision under any continuous channel distribution. 
Empirically, ChannelNet consistently matches or surpasses the performance of state-of-the-art detectors, while demonstrating superior robustness and significantly reduced computational complexity.
ChannelNet demonstrates the potential of purely data-driven detectors in massive MIMO applications. Future research could explore advanced  deep learning architectures beyond traditional MLPs and convolutional networks to further enhance performance.  
Additionally, developing few-shot adaptation strategies in new environments would be highly valuable for real-world deployment.


\appendices
\section{Proof of Theorem 1}
\textbf{Proof Sketch:}  Our goal is to show that ChannelNet-MLP serves as a universal approximator for all continuous permutation-symmetric detection functions under any continuous channel distribution. The proof assumes that the output of each antenna is one-dimensional, and it can be readily extended to high-dimensional output scenarios by separately considering each dimension. The function space of ChannelNet-MLP consists of mappings from  $\mathbb{R}^{N \times K} \times \mathbb{R}^N $ to $\mathbb{R}^{K}$. For any compact subset \(X \subseteq \mathbb{R}^{N \times K} \times \mathbb{R}^N \), we define the following function spaces:
\begin{itemize}
    \item $\mathcal{F}_{C}(X, \mathbb{R}^{K})$: the set of functions representable by ChannelNet-MLP.
    \item $\mathcal{C}_{S}(X, \mathbb{R}^{K})$: the set of all continuous permutation-symmetric functions.
\end{itemize}
We proceed in two steps:
\begin{itemize}
    \item \textbf{Approximation on a Subset $U$:} We restrict our analysis to a subset $U = \{\mathbf{(H, y)}\} \subseteq \mathbb{R}^{N \times K} \times \mathbb{R}^N$, where \(\mathbf{y}\) has distinct elements, and \(\mathbf{H}\) contains non-zero entries and has no rows or columns summing to zero. For any compact subset \(X \subseteq U \), we use the Stone-Weierstrass theorem to demonstrate that $\mathcal{F}_{C}(X, \mathbb{R}^{K})$ is dense in $\mathcal{C}_{S}(X, \mathbb{R}^{K})$. Therefore, for any \(g \in \mathcal{C}_{S}(X, \mathbb{R}^{K})\) and $\epsilon > 0$, there exists a function \(f \in \mathcal{F}_{C}(X, \mathbb{R}^{K})\) such that  \(\|f(\mathbf{H, y}) - g(\mathbf{H, y})\| < \epsilon\)  for all \((\mathbf{H, y}) \in X\).  The key here is to verify that $\mathcal{F}_{C}(X, \mathbb{R}^{K})$ satisfies the conditions of the Stone-Weierstrass theorem.
    \item \textbf{Extension to the Full Domain:} The subset $U$ has full measure with respect to any continuous probability density function on $\mathbb{R}^{N \times K} \times \mathbb{R}^N $. This means for any $\delta > 0$, there always exists a compact subset \(X \subseteq U\) with measure greater than \(1 - \delta\). From the first step, for any compact subset \(X \subseteq U \),  $\mathcal{F}_{C}(X, \mathbb{R}^{K})$ is dense in $\mathcal{C}_{S}(X, \mathbb{R}^{K})$. Since $X$ can be chosen to have measure arbitrarily close to 1, this implies that $f$ approximates $g$ within $\epsilon$ on all but a set of arbitrarily small measure,  which concludes the proof.  
\end{itemize}

\textbf{Stone-Weierstrass Theorem for Equivariant Functions:} 
In the subsequent proof, for simplicity, we assume that the antenna feature processor may take all continuous functions on given domains, following the settings in \cite{azizian2020expressive}. This assurance is provided by the universal approximation properties of MLPs \cite{cybenko1989approximation}.  The Stone-Weierstrass theorem presented below serves as the primary mathematical tool in the proof. Notably, the Stone-Weierstrass theorem employed here is tailored for equivariant functions \cite{azizian2020expressive}.  For additional details on other variations of the Stone-Weierstrass theorem, see \cite{timofte2005stone}.

Before presenting the Stone-Weierstrass theorem, we define the concept of the \textit{separation power} of a set of functions $\mathcal{F}$ on a set $X$.  The separation power is formalized through an equivalence relation, where points in $X$ that cannot be distinguished by any function in $\mathcal{F}$ are considered equivalent. The definition is as follows:

\begin{definition}
Let $\mathcal{F}$ be a set of functions defined on a set $X$.  The equivalence relation $\rho(\mathcal{F})$ induced by $\mathcal{F}$ on $X$ is defined as follows: for any $x, x' \in X$,
\begin{equation}
    (x, x' ) \in \rho (\mathcal{F})  \Leftrightarrow \forall f \in \mathcal{F}, f(x) = f(x').
\end{equation}
\end{definition}
In the case of a single function $f$, we denote the equivalence relation induced by  $f$ as $\rho (f)$, which is shorthand for  $\rho (\{f\})$. 

With the definition on separation power, the Stone–Weierstrass theorem is illustrated below.

\begin{thm}[Stone-Weierstrass Theorem \cite{azizian2020expressive}]\label{theorem:st}
Let $X$ be a compact space, $Z = \mathbb{R}^p$ for some p, $G$ be a finite group acting continuously on $X$ and $Z$, and $\mathcal{C}_E(X, Z)$ the set of all equivariant continuous functions with respect to $G$, i.e.,
\[ \mathcal{C}_E(X, Z) = \{ f \in \mathcal{C}(X, Z): f(g \cdot x) = g \cdot f(x), \forall g \in G, x\in X\},\]
where $\mathcal{C}(X, Z)$ denotes the space of all continuous functions from $X$ to $Z$. Suppose $\mathcal{F} \subseteq \mathcal{C}_E(X, Z)$ is a non-empty set of equivariant functions, and let $\pi : Z \rightarrow Z /G$ denote the canonical projection on the quotient space $Z/G$ \footnote{The quotient space $Z/G$  is formed by identifying points in $Z$ that are related by the group action $G$, effectively collapsing them into single representatives.}.  If the following conditions hold:
\begin{enumerate}
    \item $\mathcal{F}$ is a sub-algebra of $\mathcal{C}(X, Z)$ and contains the constant function $\mathbf{1}$.
    \item The set of scalar functions $\mathcal{F}_{scal} \subseteq \mathcal{C}(X, \mathbb{R})$, defined as $\mathcal{F}_{scal} = \{f \in \mathcal{C}(X, \mathbb{R}): f\mathbf{1} \in \mathcal{F}\}$, satisfies $\rho(\mathcal{F}_{scal}) \subseteq \rho(\pi \circ \mathcal{F})$, where $\pi \circ \mathcal{F}$ denotes the canonical projection $\pi$ is applied to the output of functions in the set $\mathcal{F}$, \textit{i.e.,}  $\pi \circ \mathcal{F} = \{  \pi \circ f, f\in \mathcal{F}\}$.
\end{enumerate}
Then the closure of $\mathcal{F}$ is
\begin{equation*}
    \overline{\mathcal{F}} = \{f \in \mathcal{C}_E(X, Z): \rho (\mathcal{F}) \subseteq \rho (f), \forall x \in X, f(x) \in  \mathcal{F}(x) \}
\end{equation*}
where $\mathcal{F}(x) = \{f(x), f \in \mathcal{F}\}$.
Additionally, let $I(x) = \{(i, j) \in [p]^2 : \forall \mathbf{z} \in \mathcal{F}(x), \mathbf{z}_i = \mathbf{z}_j \}$.  Then $\mathcal{F}(x) = \{ \mathbf{z} \in \mathbb{R}^p : \forall (i, j) \in I(x), \mathbf{z}_i = \mathbf{z}_j \}.$
\end{thm}


To apply the Stone-Weierstrass theorem, we define a subset $U \subseteq \mathbb{R}^{N \times K} \times \mathbb{R}^N$ as follows:
\begin{align*}
    U = \{&(\mathbf{H, y}) \; |\; \mathbf{H}_{ij} \neq 0 \; \forall i, j, \;\sum_{i=1}^N\mathbf{H}_{ij} \neq 0 \;\forall j, \\
    &\sum_{j=1}^K\mathbf{H}_{ij} \neq 0, \forall i, \; \mathbf{y}_i \neq \mathbf{y}_j, \;\forall i \neq j\}.
\end{align*}
    
We then apply the Stone-Weierstrass theorem on any compact space $X$ in $U$ to establish that $\mathcal{F}_C(X,  \mathbb{R}^{K})$ is dense in $\mathcal{C}_S(X,  \mathbb{R}^{K})$.  The subsequent lemmas are used to verify the conditions  required for applying the Stone-Weierstrass theorem.

\begin{lemma} \label{lemma:subalg}
    For any compact space $X \subset U$,  $\mathcal{F}_C(X,  \mathbb{R}^{K})$ forms a subalgebra of $\mathcal{C}(X, \mathbb{R}^K)$.
\end{lemma}

\begin{IEEEproof}[Proof of Lemma \ref{lemma:subalg}] 
To prove  that $\mathcal{F}_C(X,  \mathbb{R}^{K})$ is a subalgebra, we need to show it is closed under addition and multiplication.   
Consider any $f, \hat{f} \in \mathcal{F}_C(X,  \mathbb{R}^{K})$, where $f$  consists of $L$ iterations of antenna processors, denoted by $\{ \phi^{(1)}(\cdot),  \cdots,  \psi^{(L)}(\cdot)\}$, and $\hat{f}$ consists of $\hat{L}$ iterations of antenna processors, denoted by  $\{ \hat{\phi}^{(1)}(\cdot), \cdots, \hat{\psi}^{(\hat{L})}(\cdot)\}$.

First, we consider the case that $f$ and $\hat{f}$ have the same number of iterations, \textit{i.e.,} $L = \hat{L}$. 
We construct $f +\hat{f}  \in \mathcal{F}_C(X,  \mathbb{R}^{K})$ and $f *\hat{f}  \in \mathcal{F}_C(X,  \mathbb{R}^{K})$  as follows:
\begin{itemize}
\item The receive antenna processors in $f +\hat{f}$ and $f *\hat{f}$ are defined by:
\begin{itemize}
\item \textbf{First iteration}: Both $\phi_{+}^{(1)}(\cdot)$ and $\phi_{*}^{(1)}(\cdot)$  are constructed  by concatenating the features obtained from $\phi^{(1)}(\cdot) $ and $\hat{\phi}^{(1)}(\cdot)$:  \[\phi_{+}^{(1)}(y) = \phi_{*}^{(1)}(y) = \left(\phi^{(1)}(y), \hat{\phi}^{(1)}(y)\right),\] where $y$ represents the scalar input to the receive antenna MLP in the first iteration.  
\item \textbf{$i$-th iteration $(1 < i \leq L)$}: In subsequent iterations, both $\phi_{+}^{(i)}(\cdot)$ and $\phi_{*}^{(i)}(\cdot)$ are defined by concatenating the outputs of $\phi^{(i)}(\cdot)$ and $\hat{\phi}^{(i)}(\cdot)$ applied to their respective inputs: \[\phi_{+}^{(i)}(\mathbf{z} , \mathbf{\hat{z}}) = \phi_{*}^{(i)}(\mathbf{z} , \mathbf{\hat{z}})  = \left(\phi^{(i)}(\mathbf{z}), \hat{\phi}^{(i)}(\mathbf{\hat{z}})\right), \] 
where $\mathbf{z}$ and  $\mathbf{\hat{z}}$ represent the input features from $f$  and $\hat{f}$, respectively.
\end{itemize}
\item The transmit antenna processors in $f +\hat{f}$ and $f *\hat{f}$ are defined by:
\begin{itemize}
\item \textbf{$i$-th iteration $(1 \leq i < L)$}: Similar to the construction of   $\phi_{+}^{(i)}(\cdot)$ and $\phi_{*}^{(i)}(\cdot)$ ,  both  $\psi_{+}^{(i)}(\cdot)$ and $\psi_{*}^{(i)}(\cdot)$ are  defined by concatenating the outputs of $\psi^{(i)}(\cdot)$ and $\hat{\psi}^{(i)}(\cdot)$ applied to their respective inputs:
\[\psi_{+}^{(i)}(\mathbf{z} , \mathbf{\hat{z}}) = \psi_{*}^{(i)}(\mathbf{z} , \mathbf{\hat{z}})  = \left(\psi^{(i)}(\mathbf{z}), \hat{\psi}^{(i)}(\mathbf{\hat{z}})\right).\] 

\item \textbf{Last iteration}: $\psi_{+}^{(L)}$ is constructed by adding these two outputs while $\psi_{*}^{(L)}$ is constructed by multiplying these two outputs:
\begin{align}
    \psi_{+}^{(L)}(\mathbf{z} , \mathbf{\hat{z}}) = \psi^{(L)}(\mathbf{z} ) + \hat{\psi}^{(L)}(\mathbf{\hat{z}}), \\
    \psi_{*}^{(L)}(\mathbf{z} , \mathbf{\hat{z}}) = \psi^{(L)}(\mathbf{z} ) * \hat{\psi}^{(L)}(\mathbf{\hat{z}}). 
\end{align}

\end{itemize}
\end{itemize}
In this way, it holds
\begin{align}
    (f+\hat{f})(\mathbf{H, y}) & = f(\mathbf{H, y}) + \hat{f}(\mathbf{H, y})\\
    (f*\hat{f})(\mathbf{H, y}) & = f(\mathbf{H, y}) * \hat{f}(\mathbf{H, y}).
\end{align}
Thus, both $f + \hat{f}$ and $f * \hat{f}$ can be represented with ChannelNet-MLP.

Next, we consider the case where $L \neq \hat{L}$, and assume that $L < \hat{L}$ without loss of generality. 
We need to show the existence of an equivalent function $\tilde{f} \in \mathcal{F}_C(X,  \mathbb{R}^{K})$ with $\hat{L}$ layers, such that $\tilde{f} = f$. This reduces the problem to the first case considered.

We construct $\tilde{f}$ with $\hat{L}$ iterations of antenna processors, denoted by $\{ \tilde{\phi}^{(1)}(\cdot),  \cdots,  \tilde{\psi}^{(\hat{L})}(\cdot)\}$, as follows:
\begin{itemize}
    \item For $1 \leq i < L$: \[ \tilde{\phi}^{(i)}(\cdot) = \phi^{(i)}(\cdot), \quad \tilde{\psi}^{(i)}(\cdot) = \psi^{(i)}(\cdot).\]
    Here, $\tilde{f}$ directly adopts the receive and transmit antenna processors from $f$ for the first $L-1$ iteration.
    \item For $L \leq i < \hat{L}$: \[\tilde{\phi}^{(i)}(\cdot) = \mathbf{0}, \quad \tilde{\psi}^{(i)}(\cdot) = \psi^{(L-1)}(\cdot).\]
    Here, we set the receive antenna processor to zero in order to nullify the contribution from the current receive antenna processor. Due to the skip connections, the input to the transmit antenna processor $\tilde{\psi}^{(i)}$ remains the same as the input from the previous iteration. 
    \item For $ i = \hat{L}$: \[\tilde{\phi}^{(i)}(\cdot) = \phi^{(L)}(\cdot), \quad \tilde{\psi}^{(i)}(\cdot) = \psi^{(L)}(\cdot).\]
    In the final layer, $\tilde{f}$ adopts the receive and transmit antenna processors from the final iteration of $f$, ensuring that $\tilde{f}$ produces the same outputs with  $f$ .
\end{itemize}

This construction preserves the outputs of $f$ by ensuring that the additional layers do not alter the features until the final output. Thus substituting $f$ with $\tilde{f}$ reduces the problem to the earlier case where closure under addition and multiplication was shown. This completes the proof.
\end{IEEEproof}

The following lemma shows the separation power of $\mathcal{F}_C(X, \mathbb{R}^{K})$.



\begin{lemma}\label{lemma:f}
For any compact space $X \subset U$, $\rho\left(\mathcal{F}_C(X,  \mathbb{R}^{K})\right) = \{\left((\mathbf{H, y}) , (\mathbf{H}_{\sigma_{\mathrm{rx}}}, \mathbf{y}_{\sigma_{\mathrm{rx}}})\right), (\mathbf{H, y}) \in X, \sigma_{\mathrm{rx}} \in \mathcal{S}_N\}$, where $\mathbf{H}_{\sigma_{\mathrm{rx}}}$ represents the matrix $\mathbf{H}$ with its rows permuted by $\sigma_{\mathrm{rx}}$ and $\mathbf{y}_{\sigma_{\mathrm{rx}}}$ represents the vector $\mathbf{y}$ with its entries permuted by $\sigma_{\mathrm{rx}}$.
\end{lemma}

\begin{IEEEproof}[Proof of Lemma \ref{lemma:f}]
To prove this lemma, we will establish the equivalence between the following two statements for any $(\mathbf{H}, \mathbf{y}), (\mathbf{\hat{H}}, \mathbf{\hat{y}}) \in X$:
\begin{enumerate}
    \item For any function $f \in \mathcal{F}_C(X,  \mathbb{R}^{K})$, we have $f(\mathbf{H}, \mathbf{y}) = f(\mathbf{\hat{H}}, \mathbf{\hat{y}})$.
    \item There exists $\sigma_{\mathrm{rx}} \in \mathcal{S}_N $, such that $\mathbf{\hat{H}} = \mathbf{H}_{\sigma_{\mathrm{rx}}}$ and $\mathbf{\hat{y}} = \mathbf{y}_{\sigma_{\mathrm{rx}}}$.
\end{enumerate}

$\mathbf{2) \Rightarrow 1)}$: Assume there exists $\sigma_{\mathrm{rx}} \in \mathcal{S}_N $ such that $\mathbf{H} =\mathbf{H}_{\sigma_{\mathrm{rx}}}$ and $ \mathbf{y} = \mathbf{y}_{\sigma_{\mathrm{rx}}}$.
This permutation $\sigma_{\mathrm{rx}}$ represents a reordering of the receive antennas.  As discussed in Section III.B, ChannelNet-MLP is invariant to permutations of receive antennas. Therefore,  for any $f \in \mathcal{F}_C(X,  \mathbb{R}^{K})$, we have $f(\mathbf{H}, \mathbf{y}) = f(\mathbf{\hat{H}}, \mathbf{\hat{y}})$.

$\mathbf{1) \Rightarrow 2)}$: 
We need to show that if  no $\sigma_{\mathrm{rx}} \in \mathcal{S}_N $ exists such that  $\mathbf{H} =\mathbf{H}_{\sigma_{\mathrm{rx}}}$ and $ \mathbf{y} = \mathbf{y}_{\sigma_{\mathrm{rx}}}$, then there exists a function $f \in \mathcal{F}_C(X,  \mathbb{R}^{K})$ such that $f(\mathbf{H, y}) \neq f(\mathbf{\hat{H}, \hat{y}})$.

We consider two cases:
\begin{enumerate}
    \item \textbf{Case 1: No $\sigma_{\mathrm{rx}} \in \mathcal{S}_N$ exists such that $\mathbf{y} = \mathbf{y}_{\sigma_{\mathrm{rx}}}$}.  
     By the definition of $U$, both $\mathbf{y}$ and $\mathbf{\hat{y}}$ have distinct values. Thus, if no such $\sigma_{\mathrm{rx}}$ exists, there must be an index $i$ such that $\mathbf{y}_i \neq \mathbf{\hat{y}}_j, \forall j$. 
     Consider $f \in \mathcal{F}_C(X,  \mathbb{R}^{K})$ with a single iteration and the following antenna processors:
    \begin{itemize}
        \item $\phi(\mathbf{y}_i) = 1$,  $\phi(\mathbf{y}_j) = 0$ $\forall j \neq i$, and $\phi(\mathbf{\hat{y}}_k) = 0$ $\forall k$.
        \item $\psi(0) = 0$ and $\psi(\mathbf{H}_{ij}) = 1, \forall j$.
    \end{itemize}
    The receive antenna processor $\phi(\cdot)$ maps $\mathbf{y}_i$ to 1 and any other values in $\mathbf{y}$ and $\mathbf{\hat{y}}$ to 0 while the transmit antenna processor $\psi(\cdot)$ maps $0$ to $0$ and all non-zero entries in $\mathbf{H}$ to $1$. 
    
    Therefore, we have
    \begin{itemize}
        \item $f(\mathbf{H}, \mathbf{y}) = \mathbf{1}$: $\phi(\cdot)$ outputs $1$ for the $i$-th antenna and $0$ for the others. Consequently, after the channel layer, the input to $\psi(\cdot)$ is $\mathbf{H}_{i,j}$ for the $j$-transmit antenna.  Since $\psi(\mathbf{H}_{ij}) = 1$ $\forall j$, we have $f(\mathbf{H}, \mathbf{y}) = \mathbf{1}$.
        \item $f(\mathbf{\hat{H}}, \mathbf{\hat{y}}) = \mathbf{0}$: $\phi(\cdot)$ outputs $0$ for all antennas. After the channel layer, the input to $\psi(\cdot)$ is $0$ for all transmit antennas, leading to $f(\mathbf{\hat{H}}, \mathbf{\hat{y}}) = \mathbf{0}$.
    \end{itemize}
    Hence, $f(\mathbf{H, y}) \neq f(\mathbf{\hat{H}, \hat{y}})$, showing that $f$ can distinguish between these two inputs.
    
    \item  \textbf{Case 2: There exists $\sigma_{\mathrm{rx}}$ such that $\mathbf{y} =  \mathbf{y}_{\sigma_{\mathrm{rx}}}$,  but  $\mathbf{H} \neq \mathbf{H}_{\sigma_{\mathrm{rx}}}$}. Let $\mathbf{G} = \mathbf{H}_{\sigma_{\mathrm{rx}}}$.
    Since $\mathbf{H} \neq \mathbf{G}$,  there exists an index $i$ such that $\mathbf{H}_{i:} \neq \mathbf{G}_{i:}$.
    Consider $f \in \mathcal{F}_C(X,  \mathbb{R}^{K})$ with a single iteration and the following antenna processors:
    \begin{itemize}
        \item $\phi(\mathbf{y}_i) = 1$ and  $\phi(\mathbf{y}_j) = 0, \forall j \neq i$.
        \item $\psi(z) = z$ (identity map).
    \end{itemize}

    Thus,
    \begin{itemize}
        \item $f(\mathbf{H}, \mathbf{y}) = \mathbf{H}_{i:}$, because $\phi(\mathbf{y}_i) = 1$, activating $f$ to output the $i$-th row of $\mathbf{H}$.
        \item $f(\mathbf{\hat{H}}, \mathbf{\hat{y}}) = \mathbf{G}_{i:}$, because $\phi(\mathbf{y}_i) = 1$, activating $f$ to output the $i$-th row of $\mathbf{G}$.
    \end{itemize}
    Since $\mathbf{H}_{i:} \neq \mathbf{G}_{i:}$, $f$ can distinguish between ($\mathbf{H}, \mathbf{y})$ and $(\mathbf{\hat{H}}, \mathbf{\hat{y}})$ in this case.
\end{enumerate}
The above two cases establish that there is no $\sigma_{\mathrm{rx}} \in \mathcal{S}_N $ such that  $\mathbf{H} =\mathbf{H}_{\sigma_{\mathrm{rx}}}$ and $ \mathbf{y} = \mathbf{y}_{\sigma_{\mathrm{rx}}}$, there exists $f \in \mathcal{F}_C(X,  \mathbb{R}^{K})$ such that $f(\mathbf{H, y}) \neq f(\mathbf{\hat{H}, \hat{y}})$. This concludes the proof of Lemma \ref{lemma:f}.
\end{IEEEproof}

In addition, we consider a set of scalar functions $\mathcal{F}_{scal}$ in the following lemma.

\begin{lemma} \label{lemma:fscal}
    For any compact space $X \subset U$, there exists $\mathcal{F}_{scal}(X, \mathbb{R}) = \{f \in \mathcal{C}(X, \mathbb{R}): f\mathbf{1} \in \mathcal{F}_C(X, \mathbb{R}^K)\}$, which satisfies $\rho(\mathcal{F}_{scal}) \subseteq \rho(\pi \circ \mathcal{F}_C)$, where $\pi$ denotes the canonical projection on the quotient space $\mathbb{R}^K / \mathcal{S}_K$.
\end{lemma}

\begin{IEEEproof}[Proof of Lemma \ref{lemma:fscal}]
We define a set of scalar functions as: 
\begin{align*}
\mathcal{F}_{scal}(X, \mathbb{R}) = \{f_{scal}: & (\mathbf{H, y}) \rightarrow \max_i \left(\mathbf{H} \cdot f(\mathbf{H, y})\right)_i, \\
& f \in \mathcal{F}_C(X, \mathbb{R}^{K})\},
\end{align*}
where $\left(\mathbf{H} \cdot f(\mathbf{H, y})\right)_i$ denotes the $i$-th entry of the resulting vector from the matrix-vector product $\mathbf{H} \cdot f(\mathbf{H, y})$.
In this context, each scalar function $f_{scal}$ is derived by taking the maximum value among the components of the product vector $\mathbf{H} \cdot f(\mathbf{H, y})$, where $f$ is represented by a ChannelNet-MLP model, producing an output in $\mathbb{R}^K$.

We first demonstrate that for any $f_{scal} \in \mathcal{F}_{scal}(X, \mathbb{R})$, $f_{scal}\mathbf{1}$ can be represented by a function $\hat{f} \in \mathcal{F}_C(X, \mathbb{R}^{K})$. In other words, we can build a ChannelNet model that replicates  $f_{scal}$ across all its dimensions.

To achieve this, we construct $\hat{f}$ by extending the original function $f$ using $L+1$ iterations, where $L$ denotes the number of iterations used by $f$.
Specifically, we reuse the antenna processors from $f$ for $\hat{f}$ during the first $L$ iterations. In the $L$-th iteration (the final iteration of $f$), the transmit antenna processor $\psi^{(L)}(\cdot)$ outputs a scalar for each transmit antenna. 
This output is processed via a channel layer in $\hat{f}$ and the input to the receive antenna processor for the $L+1$ iteration becomes $\mathbf{H} \cdot f(\mathbf{H, y})$ 
The receive antenna processor in the $L+1$ iteration is defined as:
\[\phi^{(L+1)}(x) = [x, x, x^2, \cdots, x^{2N}], \]
mapping its input to a vector of length $2N+1$.

After applying $\phi^{(L+1)}(\cdot)$, the channel layer processes this vector. The input to the $j$-th transmit antenna processor is given by:
\begin{equation*} \label{equ:equations}
    \mathbf{f}^{\mathrm{tx}}_j =  [\sum_{j=1}^K \mathbf{H}_{ij}, \sum_{j=1}^K  \mathbf{H}_{ij} \mathbf{z}_j, \cdots \sum_{j=1}^K  \mathbf{H}_{ij} \mathbf{z}_j^{2N}],
\end{equation*}
where $\mathbf{z} \in \mathbb{R}^N $ represent the vector $\mathbf{H} \cdot f(\mathbf{H, y})$. Therefore $\mathbf{f}^{\mathrm{tx}}_j$ is a linear combination of vectors from the set:  \[V = \{ [1, \mathbf{z}_i, \mathbf{z}_i^2, \cdots, \mathbf{z}_i^{2N}]: i \in [N] \}.\] 
The set $V$ forms a Vandermonde matrix, implying that its vectors are linearly independent.  This independence allows us to uniquely solve for $\mathbf{z}$ using the $2N + 1$ equations provided by $ \mathbf{f}^{\mathrm{tx}}_j$ \footnote{The solution for $\mathbf{z}$ is unique. Since the vectors in $V = \{ [1, \mathbf{z}_i, \mathbf{z}_i^2, \cdots, \mathbf{z}_i^{2N}] \mid i \in [N] \}$ form a Vandermonde basis, the representation of $\mathbf{f}^{\mathrm{tx}}_j$ as their linear combination is unique. If $\mathbf{f}^{\mathrm{tx}}_j$ could also be represented by a different set of vectors $V' = \{ [1, v'_i, (v'_i)^2, \cdots, (v'_i)^{2N}] \mid i \in [N] \}$, the uniqueness of the basis implies that the sets of generating points must be identical, i.e., $\{v'_i\}_i = \{\mathbf{z}_i\}_i$.}.

The uniqueness of mapping from $\mathbf{f}^{\mathrm{tx}}_j$ to $V$ implies that we can construct an MLP model to represent this transformation.
To achieve the final output, we design $\phi^{(L+1)}(\cdot)$ to map $\mathbf{f}^{tx}_j$ to  $\max_i \mathbf{z}_i$, effectively producing the desired scalar value.

Therefore, the constructed $\hat{f} \in \mathcal{F}_C(X, \mathbb{R}^{K})$  can replicate the behavior of $f_{scal}$, thus proving that any $f_{scal} \in \mathcal{F}_{scal}$ can be represented a function $\hat{f} \in \mathcal{F}_C(X, \mathbb{R}^{K})$.

Next, we show that $\rho(\mathcal{F}_{scal}) \subseteq \rho(\pi \circ \mathcal{F}_C)$, which implies that the separate power of $\mathcal{F}_{scal}$ is stronger than $\pi \circ \mathcal{F}_C$.
According to Lemma \ref{lemma:f}, we have 
\begin{align*}
\rho\left(\mathcal{F}_C(X, \mathbb{R}^{K})\right) = \{&\left((\mathbf{H, y}) , (\mathbf{H}_{\sigma_{\mathrm{rx}}}, \mathbf{y}_{\sigma_{\mathrm{rx}}})\right): \\
& (\mathbf{H, y}) \in X, \sigma_{\mathrm{rx}} \in \mathcal{S}_N\}.
\end{align*}
Given that $\pi$ is the canonical projection associated with the transmit antenna permutation, we have:
\begin{align*}
\rho(\pi \circ \mathcal{F}_C(X, \mathbb{R}^{K})) = \{&\left((\mathbf{H, y}) , (\mathbf{H}_{\sigma_{\mathrm{tx}}, \sigma_{\mathrm{rx}}}, \mathbf{y}_{\sigma_{\mathrm{rx}}})\right): \\
& (\mathbf{H}, \mathbf{y}) \in X, \sigma_{\mathrm{rx}} \in \mathcal{S}_N, \sigma_{\mathrm{tx}} \in \mathcal{S}_K.\}
\end{align*}
To show that $\rho(\mathcal{F}_{scal}) \subseteq \rho(\pi \circ \mathcal{F}_C)$, we must show that if $f(\mathbf{H}, \mathbf{y}) = f(\mathbf{\hat{H}}, \mathbf{\hat{y}})$ holds for all $f \in \mathcal{F}_{scal}$,  there exists permutations $\sigma_{\mathrm{rx}} \in \mathcal{S}_N$ and $\sigma_{\mathrm{tx}} \in \mathcal{S}_K$ such that $(\mathbf{H}, \mathbf{y}) = (\mathbf{\hat{H}}_{\sigma_{\mathrm{tx}}, \sigma_{\mathrm{rx}}}, \mathbf{\hat{y}}_{\sigma_{\mathrm{rx}}})$. 

We will prove this by contradiction. Assume there do not exist such permutations.  We consider the following two cases:
 \begin{enumerate}
     \item \textbf{No receive permutation $\sigma_{\mathrm{rx}}$ exists:} If no $\sigma_{\mathrm{rx}}$ exists such that $\mathbf{y} = \mathbf{\hat{y}}_{\sigma_{\mathrm{rx}}}$, from the proof of Lemma \ref{lemma:f}, we can find a function $f \in \mathcal{F}_C(X,  \mathbb{R}^{K})$ such that $f(\mathbf{H}, \mathbf{y}) = \mathbf{0}$ while $f(\mathbf{\hat{H}}, \mathbf{\hat{y}}) = \mathbf{1}$. Considering the scalar function $f_{scal}$ corresponding to $f$, we have $f_{scal} (\mathbf{H, y}) = \max_i \left(\mathbf{H} \cdot f(\mathbf{H, y})\right)_i =\max_i \left(\mathbf{H} \cdot \mathbf{0}\right)_i =0$, while $f_{scal} (\mathbf{\hat{H}, \hat{y}})  = \max_i \left(\mathbf{H} \cdot \mathbf{1}\right)_i \neq 0$.
     
    \item \textbf{No transmit permutation $\sigma_{\mathrm{tx}}$ exists:}  Suppose  $\sigma_{\mathrm{rx}}$ exists such that $\mathbf{y} = \mathbf{\hat{y}}_{\sigma_{\mathrm{rx}}}$, but no $\sigma_{\mathrm{tx}}$ satisfies $\mathbf{H} = \mathbf{\hat{H}}_{\sigma_{\mathrm{tx}}, \sigma_{\mathrm{rx}}}$.  Let $\mathbf{G} = \mathbf{H}_{\sigma_{\mathrm{rx}}}$.   In this case, there exist $\mathbf{H}_{:i}$ from $\mathbf{H}$ and $\mathbf{G}_{:j}$ from $\mathbf{G}$ that appear a different number of times in $\mathbf{H}$ and $\mathbf{G}$. Denote these frequencies as follows:
    \begin{itemize}
        \item  $k_1$ and $k_2$ denote the frequency of $\mathbf{H}_{:i}$ in $\mathbf{H}$ and $\mathbf{G}$, respectively.
        \item $k'_1$ and $k'_2$ denote the frequency of  $\mathbf{G}_{:j}$  in $\mathbf{H}$ and $\mathbf{G}$, respectively.
    \end{itemize}
   
     By definition, $k_1\neq k'_1$ and $k_2\neq k'_2$. Now, consider $f\in\mathcal{F}_C(X,  \mathbb{R}^{K})$ with a single iteration using the following definitions for $\phi(\cdot)$ and $\psi(\cdot)$:
    \begin{itemize}
        \item \textbf{Receive antenna processor $\phi(\cdot)$}: The receive antenna processor $\phi(\cdot)$ maps $\mathbf{y}_i$ to a one-hot vector $\mathbf{e}_i$ of length $N$: \[\phi(\mathbf{y}_i) = \mathbf{e}_i,\] where $\mathbf{e}_i = [0, 0, \cdots, 1, \cdots ,0]$ with only the $i$-th entry being 1. 
        
        \item \textbf{Transmit antenna processor $\psi(\cdot)$}: $\psi(\cdot)$ assigns value as follows:
        \begin{align*}
        \psi(\mathbf{H}_{:i}) &= c_1,  \psi(\mathbf{G}_{:j}) = c_2, \\
        \psi(\mathbf{H}_{:k}) &= 0, \quad \text{for all other columns.}
        \end{align*}
    \end{itemize}
    Using these definitions, consider the scalar function $f_{scal}$ corresponding to $f$, we have 
    \begin{align*}
    f_{scal} (\mathbf{H, y}) &= \max_i \left(\mathbf{H} f(\mathbf{H, y})\right)_i \\
    &= \max_i \left(c_1k_1 \mathbf{H}_{:i} + c_2k_2 \mathbf{G}_{:j}\right),
    \end{align*}
    while 
    \begin{align*}
    f_{scal} (\mathbf{\hat{H}, \hat{y}}) &= \max_i \left(\mathbf{\hat{H}} f(\mathbf{\hat{H}, \hat{y}})\right)_i \\
    &= \max_i \left(c_1k'_1 \mathbf{H}_{:i} + c_2k'_2 \mathbf{G}_{:j}\right).
    \end{align*}
Since $k_1\neq k'_1$ and $k_2\neq k'_2$ and $c_1$ and $c_2$ can be arbitrarily chosen, we can select values  $c_1$ and $c_2$ such that $f_{scal} (\mathbf{H, y}) \neq f_{scal} (\mathbf{\hat{H}, \hat{y}})$
 \end{enumerate}
The above two cases establish that if no  $\sigma_{\mathrm{rx}}$ and $\sigma_{\mathrm{tx}}$ exist such that $(\mathbf{H}, \mathbf{y}) = (\mathbf{\hat{H}}_{\sigma_{\mathrm{tx}}, \sigma_{\mathrm{rx}}}, \mathbf{\hat{y}}_{\sigma_{\mathrm{rx}}})$, then there exists $f_{scal} \in \mathcal{F}_{scal}$, such that $f_{scal}(\mathbf{H}, \mathbf{y}) \neq f_{scal}(\mathbf{\hat{H}}, \mathbf{\hat{y}})$, which concludes the proof of Lemma \ref{lemma:fscal}.
\end{IEEEproof}

Based on the preceding lemmas, we now establish a key lemma that characterizes the expressive power of $\mathcal{F}_C(X,  \mathbb{R}^{K})$.
\begin{lemma}\label{lemma:dense}
For any compact space $X \subseteq U$, $\overline{\mathcal{F}_C(X,  \mathbb{R}^{K})} = \mathcal{C}_{S}(X,  \mathbb{R}^{K})$.
\end{lemma}

\begin{IEEEproof}[Proof of Lemma \ref{lemma:dense}]
 To prove this lemma, we utilize the Stone-Weierstrass theorem. Let $G$ be the permutation group $\mathcal{S}_K$ and let $Z=\mathbb{R}^K$. The conditions of the Stone-Weierstrass theorem are established as follows:
\begin{itemize}
    \item Condition 1 follows directly from Lemma \ref{lemma:subalg},  which shows that $\mathcal{F}_C(X,  \mathbb{R}^{K})$ is a sub-algebra. Moreover, the constant function $\mathbf{1}$ is in $\mathcal{F}_C(X,  \mathbb{R}^{K})$, as it can be realized using a single-iteration ChannelNet-MLP where the transmit antenna processor $\psi(\cdot)$ maps any input to $1$. 
    \item Condition 2 directly follows from Lemma \ref{lemma:fscal}.
\end{itemize}
Therefore, by applying the Stone-Weierstrass theorem, we conclude that the closure of $\mathcal{F}_C(X,  \mathbb{R}^{K})$ is given by
\begin{align*}
     \overline{\mathcal{F}_C(X,  \mathbb{R}^{K})} = \{ & f \in \mathcal{C}_E(X, \mathbb{R}^K): \rho \left(\mathcal{F}_C(X,  \mathbb{R}^{K})\right) \subseteq \rho (f), \\
                                         & \forall (\mathbf{H, y}) \in X, f(\mathbf{H, y}) \in  \mathcal{F}_C(\mathbf{H, y}) \},
\end{align*}
where $\mathcal{F}_C(\mathbf{H, y}) = \{f(\mathbf{H, y}), f \in \mathcal{F}_C\} = \{ \mathbf{z} \in \mathbb{R}^K : \forall (i, j) \in I(\mathbf{H, y}), \mathbf{z}_i = \mathbf{z}_j\}$ and $I(\mathbf{H, y}) = \{(i, j) \in [K]^2 : \forall f \in \mathcal{F}_C(X,  \mathbb{R}^{K}), f(\mathbf{H, y})_i = f(\mathbf{H, y})_j \}$. 

To complete the proof, we need to remove the conditions and show that $ \overline{\mathcal{F}_C(X,  \mathbb{R}^{K})} = \mathcal{C}_{S}(X,  \mathbb{R}^{K})$. This involves establishing the following two statements:
\begin{itemize}
    \item (1) For any $f \in  \mathcal{C}_S(X, \mathbb{R}^K)$, we have $\rho \left(\mathcal{F}_C(X,  \mathbb{R}^{K})\right) \subseteq \rho (f)$.
    \item (2) For any $f \in  \mathcal{C}_S(X, \mathbb{R}^K)$ and any $({\mathbf{H}, \mathbf{y}}) \in X$, it holds that $f({\mathbf{H}, \mathbf{y}})_i = f({\mathbf{H}, \mathbf{y}})_j$, $\forall (i,j) \in I({\mathbf{H}, \mathbf{y}})$, where 
    \begin{align*}
    I({\mathbf{H}, \mathbf{y}}) = \{ &(i, j) \in [K]^2 : \\
    & g({\mathbf{H}, \mathbf{y}})_i = g({\mathbf{H}, \mathbf{y}})_j, \; \forall g\in  \mathcal{F}_C(X,  \mathbb{R}^{K}).\}
    \end{align*}
\end{itemize}

The first statement can be readily proven using Lemma \ref{lemma:f}, which demonstrates  that $\rho \left(\mathcal{F}_C(X,  \mathbb{R}^{K})\right) = \{\left( (\mathbf{H, y}), (\mathbf{H}_{\sigma_{\mathrm{rx}}}, \mathbf{y}_{\sigma_{\mathrm{rx}}})\right), \sigma_{\mathrm{rx}} \in \mathcal{S}_N\}$.
According to the definition of permutation-symmetric detection function, for every $f \in  \mathcal{C}_S(X, \mathbb{R}^K)$, it holds that $f(\mathbf{H, y}) = f(\mathbf{H}_{\sigma_{\mathrm{rx}}}, \mathbf{y}_{\sigma_{\mathrm{rx}}})$, therefore, $\rho \left(\mathcal{F}_C(X, \mathbb{R}^K)\right) \subseteq \rho(f)$.

For the second statement, consider a function set consisting of $N$ functions $\mathcal{F}_N = \{f_n \in \mathcal{F}_C(X, \mathbb{R}^K), n \in [N]\}$, where each function $f_n$ is a single iteration ChannelNet-MLP with antenna feature processors $\phi_n(\cdot)$ and $\psi_n(\cdot)$ defined as follows:
\begin{itemize}
    \item $\phi_n(\mathbf{y}_n) = 1$, $\phi_n(\mathbf{y}_i) = 0, \forall i \neq n$.
    \item $\psi_n(z) = z$. 
\end{itemize}
Similar to the example in the proof for Lemma \ref{lemma:f}, $f_n$  outputs the $n$th row of $\mathbf{H}$, \textit{i.e.,} $f_n(\mathbf{H, y}) = \mathbf{H}_{n:}$. 
Let $I_N(\mathbf{H, y})$ denote the set of index pairs $\{(i,j) \in [K]^2 , f_n(\mathbf{H, y})_i = f_n(\mathbf{H, y})_j, \forall f_n\in \mathcal{F}_N \}$. 
If $(i, j) \in I_N(\mathbf{H, y})$, it implies that $\mathbf{H}_{:i} = \mathbf{H}_{:j}$.
Therefore, there exists a permutation $\sigma_{\mathrm{tx}} \in \mathcal{S}_K$ that switches $i$ and $j$, such that $\mathbf{H} = \sigma_{\mathrm{tx}}(\mathbf{H})$.
Therefore, for any $f \in \mathcal{C}_S(X, \mathbb{R}^K)$,  we have $f(\mathbf{H, y})_i = \sigma_{\mathrm{tx}} \left(f(\mathbf{H, y})\right)_j = f(\sigma_{\mathrm{tx}}(\mathbf{H}), \mathbf{y})_j = f(\mathbf{H, y})_j$. 
Since $\mathcal{F}_N \subseteq \mathcal{F}_C(X, \mathbb{R}^K)$, it follows that $I(\mathbf{H, y}) \subseteq I_N(\mathbf{H, y})$.
Therefore, for any $(i, j) \in I(\mathbf{H, y})$, $f(\mathbf{H, y})_i = f(\mathbf{H, y})_j$.

With these two statements proven, we conclude that $\overline{\mathcal{F}_C(X, \mathbb{R}^K)} = \mathcal{C}_{S}(X, \mathbb{R}^K)$.
\end{IEEEproof}

\begin{IEEEproof}[Proof of Theorem 1]
Let \( p(\mathbf{H, y}) \) denote any continuous probability density function on \( \mathbb{R}^{N \times K} \times \mathbb{R}^N \). The complement of \( U \), denoted as \( U^c \), consists of degenerate subspaces, which have measure zero under any continuous \( p(\mathbf{H, y}) \). Therefore, the measure of \( U \) is one, \textit{i.e.}, \( \mathbb{P}(U) = 1 \).

Given any \( \delta > 0 \), we can select a compact subset \( X \subset U \) such that \( \mathbb{P}(X) > 1 - \delta \). By applying Lemma \ref{lemma:dense}, we know that  \( \mathcal{F}_C(X, \mathbb{R}^K) \) is dense in  \( \mathcal{C}_S(X, \mathbb{R}^K) \). Specifically, for any \( g \in \mathcal{C}_S(X, \mathbb{R}^K) \) and any \( \epsilon > 0 \), there exists a function \( f \in \mathcal{F}_C(X, \mathbb{R}^K) \) such that:\[ \| f(\mathbf{H, y}) - g(\mathbf{H, y}) \| < \epsilon, \forall (\mathbf{H, y}) \in X. \]
Since \( \mathbb{P}(X) > 1 - \delta \), the approximation error satisfies: \[ \mathbb{P}( \| f(\mathbf{H, y}) - g(\mathbf{H, y}) \| < \epsilon ) > 1 - \delta, \]
which concludes the proof.

\end{IEEEproof}

\section{Proof of Corollary}

To prove this corollary,  we first demonstrate that the ML detector $f_{ML}$ is permutation-symmetric and measurable.  
Following that, we show while $f_{ML}$ is not continuous, its measurability is sufficient for proving the corollary.
\begin{lemma}\label{lemma:ml}
    $f_{ML}$ is measurable and permutation-symmetric.
\end{lemma}

\begin{IEEEproof}[Proof of Lemma \ref{lemma:ml}]
The ML detector $f_{ML}$ can be expressed as \[f_{ML}(\mathbf{H}, \mathbf{y}) = \arg\min_{\mathbf{x} \in \mathcal{X}^K}\left(d_{(\mathbf{H}, \mathbf{y})}(\mathbf{x})\right),\] where $d_{(\mathbf{H}, \mathbf{y})}(\mathbf{x}) = \| \mathbf{Hx} - \mathbf{y} \|_2^2$.

\textbf{Permutation-Symmetry:} For any $\sigma_{\mathrm{rx}} \in \mathcal{S}_N$ with a corresponding permutation matrix $\mathbf{P}_{\mathrm{rx}} \in \{0, 1\}^{N \times N}$, we have
\begin{align*}
    d_{\left(\mathbf{H}_{\sigma_{\mathrm{rx}}},\mathbf{y}_{\sigma_{\mathrm{rx}}}\right)}(\mathbf{x})  & = \|\mathbf{P}_{\mathrm{rx}} \mathbf{H}  \mathbf{x} - \mathbf{P}_{\mathrm{rx}} \mathbf{y}\|_2^2  = \|\mathbf{P}_{\mathrm{rx}}(\mathbf{Hx-y})\|_2^2. 
\end{align*} 
Since $\|\mathbf{P}_{\mathrm{rx}} \mathbf{v}\|_2 = \|\mathbf{v}\|_2$ holds for any vector $ \mathbf{v}$, we have $ d_{\left(\mathbf{H}_{\sigma_{\mathrm{rx}}},\mathbf{y}_{\sigma_{\mathrm{rx}}}\right)}(\mathbf{x}) = d_{(\mathbf{H}, \mathbf{y})}(\mathbf{x})$. Therefore, $f_{ML}(\mathbf{H}, \mathbf{y}) = f_{ML}\left(\mathbf{H}_{\sigma_{\mathrm{rx}}},\mathbf{y}_{\sigma_{\mathrm{rx}}}\right)$.

Next, consider any permutation $\sigma_{\mathrm{tx}}\in \mathcal{S}_K$ with the corresponding permutation matrix $\mathbf{P}_{\mathrm{tx}} \in \{0, 1\}^{K\times K}$.
We have 
\begin{align*}
    d_{\left(\mathbf{H}_{\sigma_{\mathrm{tx}}}, \mathbf{y}\right)}(\mathbf{x}) &= \|\mathbf{H} \mathbf{P}_{\mathrm{tx}} \mathbf{x} - \mathbf{y}\|_2^2 = d_{\left(\mathbf{H}, \mathbf{y}\right)}(\mathbf{x}_{\sigma_{\mathrm{tx}}}).
\end{align*}
Hence, $f_{ML}(\mathbf{H}_{\sigma_{\mathrm{tx}}}, \mathbf{y}) = \sigma_{\mathrm{tx}}(f_{ML}(\mathbf{H}, \mathbf{y}))$. Therefore, we can conclude $f_{ML}$ is permutation-symmetric, satisfying $\sigma_{\mathrm{tx}}(f_{ML}(\mathbf{H}, \mathbf{y})) = f_{ML}\left(\mathbf{H}_{\sigma_{\mathrm{rx}}, \sigma_{\mathrm{tx}}}, \mathbf{y}_{\sigma_{\mathrm{rx}}}\right)$.

\textbf{Measurability:} To demonstrate that $f_{ML}$ is measurable, consider the preimage of $f_{ML}$ for each point in $\mathcal{X}^K$.
For any $\mathbf{x} \in \mathcal{X}^K$, consider the function $g_{\mathbf{x}}$ given by
\begin{equation}
    g_{\mathbf{x}}(\mathbf{H}, \mathbf{y}) = \| \mathbf{Hx}-\mathbf{y}\|_2^2 - \min_{\mathbf{z\in\mathcal{X}^K}} \| \mathbf{Hz}-\mathbf{y}\|_2^2.
\end{equation} 
The preimage $f_{ML}^{-1}(\mathbf{x})$ can be expressed as $g_{\mathbf{x}}^{-1}(0)$.
The function $g_{\mathbf{x}}$ is continuous and therefore measurable due to the continuity of the squared norm. Consequently, since $g_{\mathbf{x}}^{-1}(0)$ is measurable, it follows that $f_{ML}^{-1}(\mathbf{x})$ is measurable for each point in $\mathcal{X}^K$, proving that $f_{ML}$ is  measurable.
\end{IEEEproof}

With the following Lusin’s theorem, the measurability of $f_{ML}$ is sufficient for the proving the Corollary. 

\begin{thm}[Lusin’s theorem \cite{evans2018measure}]
Let $f: X \rightarrow \mathbb{R}$ be a measurable function defined on a Lebesgue measurable set $X \subseteq \mathbb{R}^p$ for which the Lebesgue measure $l(X)$ is finite. For any $\delta > 0$ there exists a compact subset $ W \subseteq X$ such that $l(X \backslash W) < \delta$ and $f$ is continuous on $W$.
\end{thm}

\begin{IEEEproof}[Proof of Corollary]
From Lemma \ref{lemma:ml}, $f_{ML}$ is measurable and permutation-symmetric.  
Let \( p(\mathbf{H, y}) \) denote any continuous probability density function on \( \mathbb{R}^{N \times K} \times \mathbb{R}^N \).
According to Lusin’s theorem, for any $\delta>0$, there exists a compact subset $X \subseteq U$ such that $\mathbb{P}(X) > 1- \delta$ and $f_{ML}$ is continuous on $X$. 
Applying Lemma \ref{lemma:dense}, for any $\epsilon>0$,  there exists  $f \in  \mathcal{F}_C(X, \mathbb{R}^K)$ such that $\|f(\mathbf{H, y})-f_{ML}(\mathbf{H, y})\| < \epsilon, \forall (\mathbf{H, y}) \in X$.
Since $\mathbb{P}(X) > 1 - \delta$, it follows that $\mathbb{P}(\|f(\mathbf{H, y})-f_{ML}(\mathbf{H, y})\| < \epsilon) > 1-\delta$, which concludes the proof.
\end{IEEEproof}

\section{Network architecture}
\begin{table}[h!]
\centering
\renewcommand\theadalign{cc}
\renewcommand\cellalign{cc}
{\begin{tabular}{cccc}
\toprule
\makecell{\textbf{Layers}} & 
\makecell{\textbf{Kernel}\\ \textbf{size}} & 
\makecell{\textbf{Number of }\\ \textbf{filters/neurons}} & 
\makecell{\textbf{Used in}} \\
\midrule
\multirow{2}{*}{\makecell{CONV layer 1\\CONV layer 2}} & 
\multirow{2}{*}{\makecell{3\\3}} & 
\multirow{2}{*}{\makecell{$d$\\$d$}} & 
\multirow{2}{*}{ChannelNet-Conv only} \\
 &  &  &  \\
\midrule
\multirow{2}{*}{\makecell{MLP layer 1\\MLP layer 2}} & 
\multirow{2}{*}{\makecell{$-$\\$-$}} & 
\multirow{2}{*}{\makecell{$d$\\$d$}} & 
\multirow{2}{*}{\makecell{\(\phi^{(t)}\) (\(t=1,\dots,L\)) and \\ \(\psi^{(t)}\) (\(t=1,\dots,L-1\)) }} \\
 &  &  &  \\
\midrule
\multirow{3}{*}{\makecell{MLP layer 1\\MLP layer 2\\MLP layer 3}} & 
\multirow{3}{*}{\makecell{$-$\\$-$\\$-$}} & 
\multirow{3}{*}{\makecell{\(5d\)\\\(5d\)\\\(M\)}} & 
\multirow{3}{*}{\(\psi^{(L)}\) } \\
 &  &  &  \\
 &  &  &  \\
\bottomrule
\end{tabular}}
\caption{Network architecture for ChannelNet-MLP and ChannelNet-Conv. The ``Used in'' column indicates the module in which each layer is applied.}
\end{table}

\bibliographystyle{IEEEtran}
\bibliography{reference}

\end{document}